\documentclass[pra,twocolumn,superscriptaddress,notitlepage]{revtex4-1}
\usepackage{graphicx,amsmath,amsfonts,amssymb,upgreek,txfonts,color}
\usepackage[colorlinks,linkcolor=blue,citecolor=red,urlcolor=blue,breaklinks=true]{hyperref}
\begin{document}
	
\title{Dynamical Casimir effect of phonon excitation in the dispersive regime of cavity optomechanics}

\author{Ali Motazedifard} 
\email{motazedifard.ali@gmail.com}
\address{ Department of Physics, Faculty of Science, University of Isfahan, Hezar Jerib, 81746-73441, Isfahan, Iran}

\author{M. H. Naderi} 
\email{mhnaderi@phys.ui.ac.ir}
\address{ Department of Physics, Faculty of Science, University of Isfahan, Hezar Jerib, 81746-73441, Isfahan, Iran}
\address{Quantum Optics Group, Department of Physics, Faculty of Science, University of Isfahan, Hezar Jerib, 81746-73441, Isfahan, Iran}

\author{R. Roknizadeh}
\email{rokni@sci.ui.ac.ir}
\address{ Department of Physics, Faculty of Science, University of Isfahan, Hezar Jerib, 81746-73441, Isfahan, Iran}
\address{Quantum Optics Group, Department of Physics, Faculty of Science, University of Isfahan, Hezar Jerib, 81746-73441, Isfahan, Iran}

\date{\today}
\begin{abstract}
	In this paper, we theoretically propose and investigate a feasible experimental scheme for realizing the dynamical Casimir effect (DCE) of phonons in an optomechanical setup formed by a ground-state precooled mechanical oscillator (MO) inside a Fabry-P{\'e}rot cavity, which is driven by an amplitude-modulated classical laser field in the dispersive (far-detuned) regime. The time modulation of the driving field leads to the parametric amplification of the mechanical vacuum fluctuations of the MO, which results in the generation of Casimir phonons over time scales longer than the cavity lifetime. We show that the generated phonons exhibit quadrature squeezing, bunching effect, and super-Poissonian statistics which are controllable by the externally modulated laser pump. In particular, we find that the scheme allows for a perfect squeezing transfer from one mechanical quadrature to another when the laser frequency is varied from red detuning to blue detuning. Moreover, by analyzing the effect of the thermal noise of the MO environment, we find that there exists a critical temperature above which there is no phonon quadrature squeezing to occur. We also show that in the presence of time modulation of the driving laser the linewidth narrowing of the displacement spectrum of the MO can be considered as a signature of the generation of Casimir phonons.
\end{abstract}

\pacs{, , , }

\maketitle

\section{Introduction}
A distinctive feature of the quantum theory of the electromagnetic field, and of quantum field theory in general, consists of the existence of quantum vacuum fluctuations which have no counterpart in the classical world. One of the manifestations of such quantum fluctuations in macroscopic level is the generation of real particles out of the vacuum when the boundary conditions of the field are varied in time at a fast-enough rate\cite{Moore,Davies,Yablonovitch,Dodonov1,Nation}, an effect known as the \textit{dynamical} Casimir effect (DCE). This effect which arises from a mismatch of vacuum modes in time domain, can be explained qualitatively as a particular case of the parametric amplification of vacuum fluctuations in the systems with time-dependent parameters \cite{Yablonovitch,Schwinger,Wilson}. The possibility of particle creation via the DCE has been theoretically identified and investigated in a large variety of systems. Examples range from cosmology, such as particle creation as a consequence of an expanding gravitational background \cite{Davies book}, to non-stationary cavity QED, such as photon generation in Fabry-P{\'e}rot cavities with moving mirrors \cite{Moore,Dodonov2,Dalvit,Dodonov3,Crocce,Dodonov4,Dodonov5}. Moreover, various theoretical schemes for practical applications of the DCE have been proposed, including generation of photons with nonclassical properties \cite{Dodonov6,Dodonov7,Johansson}, generation of atomic squeezed sates \cite{Bhattacherjee}, generation of multipartite entanglement in cavity networks \cite{Felicetti}, and generation of EPR quantum steering and Gaussian interferometric power \cite{Sabin}.

Although motion-induced DCE has been theoretically studied for more than four decades, the effect has not yet been demonstrated by experiments directly. The main difficulty is due to the fact that for a measurable flux of real photons to be generated, the moving boundaries should oscillate at very high frequencies that are not yet experimentally accessible. Consequently, alternative schemes based on imitation of boundary motion have been proposed. For example, the boundary motion may be replaced by suitable periodic modulation of the optical properties of the boundary \cite{Lombardi,Dodonov8,Dodonov9} or of the optical path length of a cavity \cite{Dezael,Faccio,Motazedifard DCE}. Some other experimental schemes aiming the observation of the DCE can be found in \cite{Lombardi,Agnesi,Kawakubo}. Recently, it has been reported about a successful implementation of DCE in superconducting circuits through two independent experiments by fast-modulating either the electrical boundary condition of a transmission line \cite{Pourkabirian} or the effective speed of light in a Josephson metamaterial \cite{Lahteenmaki}.

Besides the DCE for photons, the dynamical Casimir emission of particles other than photons have also been investigated in recent years. Among the proposed schemes in this direction we can cite, for example, the DCE of phonons in a time modulated atomic Bose-Einstein condensate (BEC) \cite{Recati,Jaskula}, dynamical Casimir emission of Bogoliubov excitations in an exciton-polariton condensate suddenly created by an ultrashort laser pulse \cite{Koghee}, phononic DCE in a time-modulated quantum fluid of light \cite{Busch}, DCE of magnon excitation in a
spinor BEC driven by a time-dependent magnetic field \cite{Saito}, DCE of phonons in a gas of laser-cooled atoms with time-dependent effective charge \cite{Dodonov10}, and the DCE of phonons in a non-stationary quantum well-assisted optomechanical cavity driven by an amplitude-modulated external laser pump \cite{Mahajan}.

In the present paper, we propose and analyze a feasible experimental scheme to realize the DCE of phonon excitation in the so-called membrane-in-the-middle optomechanical system \cite{Thompson,Jayich} which is pumped by a far-detuned driving classical laser and also precooled down to near its ground state by an auxiliary cavity mode. We show that the time-modulation of the input power of the pump laser leads to the parametric amplification of the mechanical vacuum fluctuations of the membrane resulting in pair phonons (Casimir phonons) creation over time scales longer than the cavity lifetime. By investigating the quantum statistical properties of the generated Casimir phonons in both red- and blue- detuned regimes, we find that they exhibit motional quadrature squeezing, bunching effect, and super-Poissonian statistics, which are controllable through adjusting the parameters of the driving laser. In particular, the results reveal that whereas the phonon counting statistics is the same in both red- and blue- detuned regimes, the motional squeezing can completely be transferred from one mechanical quadrature to another when the laser frequency is varied from red detuning to blue detuning. Moreover, phonon quadrature squeezing is sensitive to the initial mean number of thermal phonons of the MO such that for temperatures higher than a critical value the generated phonons exhibit no squeezing, that is why the MO should be precooled.

The paper is structured as follows. In section \ref{sec2}, we describe the system under consideration and derive an effective system Hamiltonian in dispersive regime. We also calculate the mean number of generated Casimir phonons and examine its temporal behavior within as well as beyond the rotating wave approximation (RWA). In section \ref{sec3}, we study the quantum statistical properties of the generated Casimir phonons, including mechanical quadrature squeezing and phonon counting statistics. In section \ref{sec4}, we show that the generation of the Casimir phonons leads to the linewidth narrowing of the displacement spectrum of the MO. Finally, our conclusions are summarized in section \ref{summary}.

\section{description of the system and DCE of phonons } \label{sec2}
\begin{figure}
	\includegraphics[width=8.5cm]{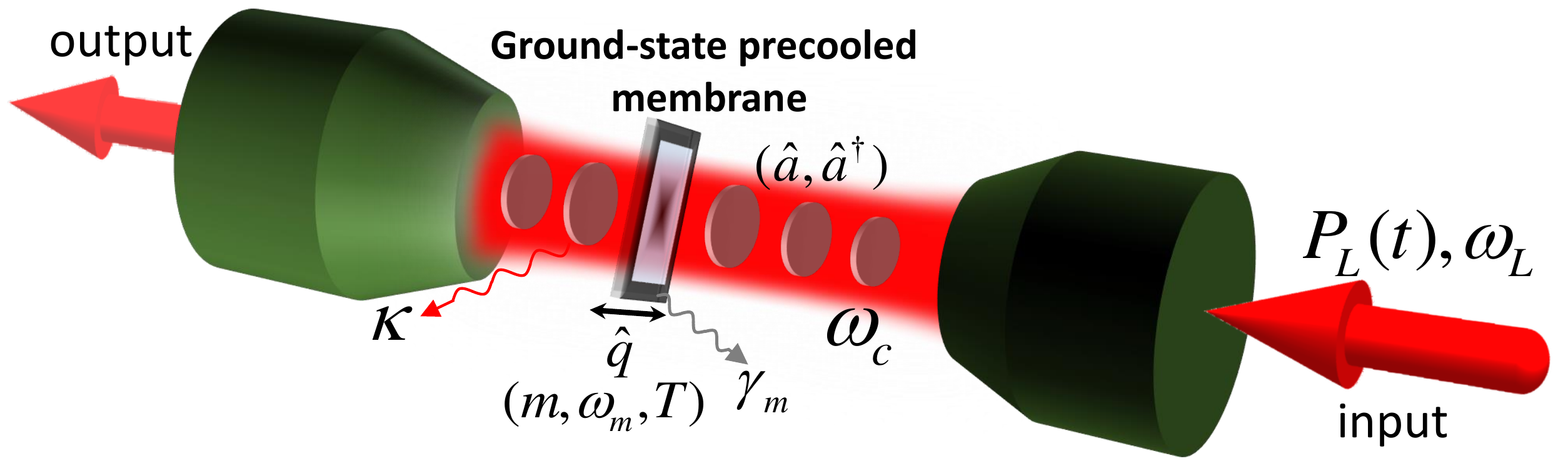}
	\caption{( Color online) Schematic description of the system under consideration. A mechanical oscillator (MO) with effective mass $m$ and frequency $\omega_m$ is placed inside a single-mode Fabry-P\'{e}rot cavity with length $L$ and frequency $\omega_c$. The cavity is first pumped by a  far-detuned classical laser field of frequency $\omega_L$ and constant power $P_L$ until the cavity mode can adiabatically be eliminated. Then, the cavity is pumped by a driving field with harmonically modulated power $P_L(t)$. Dissipations occur via cavity decay and mechanical damping with rates $\kappa$ and $\gamma_m$, respectively. Moreover, the MO is first cryogenically precooled and then cooled down to near its ground state by an auxiliary cavity mode in the resolved sideband and weak coupling regimes.}
	\label{Fig1}
\end{figure}

\subsection{The system}

As shown in Fig. (\ref{Fig1}), we consider a membrane-in-the-middle optomechanical system in which a thin dielectric membrane is placed inside a high-finesse Fabry-P\'{e}rot cavity with length $L$. We also assume that the membrane is precooled into its ground state. The membrane can be modelled as a single-mode MO of intrinsic frequency $\omega_m$, effective mass $m$, and damping rate $\gamma_m$. In the proposed scheme, we assume that a single cavity mode of natural frequency $\omega_c=\pi c / L$ (in the absence of the middle membrane) and decay rate $\kappa$ is driven by a far-detuned classical laser field of frequency $\omega_L$ and constant power $P_L$ until the cavity mode can adiabatically be eliminated. Then, the modulation of the driving field is suddenly switched on according to $P_L(t)=P_L(1+\varepsilon \cos(\Omega t))$ where $\Omega$ and $\varepsilon$ ($<1 $) are the frequency and amplitude of the modulation, respectively. 
Recently, it has been shown \cite{Mari1,Giovannetti,Mari2} that parametrically modulated optomechanical systems can be exploited to achieve large degrees of mechanical squeezing as well as to enhance the efficiency of photon-phonon entanglement. Very recently, in Ref. \cite{two-phonon driving clerk}, the authors have used modulation technique to implement coherent two-phonon driving, i.e., the degenerate parametric amplification (DPA) of the mechanical mode in an optomechanical cavity. They have demonstrated how this system can be used to perform single-quadrature detection of a near-resonant narrow-band force sensing with extremely low added noise. Moreover, the cavity spectral function in their proposed scheme can become negative under certain condition which is strongly different from a conventional DPA and is responsible for the optomechanical induced transparency (OMIT). Regarding to the experimental investigation, it has been very recently reported \cite{PRL modulation} the observation 
of two-mode squeezing in the oscillation quadratures of a thermal micro-oscillator by exploiting parametric modulation of the optical spring in an optomechanical cavity. 

In the model under consideration, the single-cavity and mechanical modes approximation is an appropriate simplification provided that the cavity free spectral range is much larger than the mechanical frequency \cite{singlecavitymoderegime} and the detection bandwidth is chosen such that it includes only a single, isolated, mechanical resonance and mode-mode coupling is negligible\cite{singlemechanicalmoderegime}. Furthermore, in this type of optomechanical system the frequency of cavity depends on the MO displacement, $q$, i.e., ${\omega _c}(q) =\omega_c+ (c/L ){\cos ^{ - 1}}\left( {\left| {{r_c}} \right|\cos (4\pi q/{\lambda _c})} \right)$ \cite{Jayich} where $r_c$ and $\lambda_c$ are, respectively, the reflectivity of the MO and the wavelength of the cavity mode, and the position of the MO is calculated from the anti-node of the cavity field. It has been shown \cite{Sh. Barzanjeh} that this dependence leads to a nonlinear coupling (phonon number-dependent optomechanical coupling) of the radiation pressure field with the MO through multi-phonon excitations of the vibrational sidebands. However, by considering the first excitation of the vibrational sideband in the limit of very small values of the Lamb-Dicke parameter, $\eta  = (4\pi /{\lambda _c})\sqrt {\pi \hbar /(m{\omega _m})}$, and for low values of the membrane reflectivity, the phonon number dependence of the optomechanical coupling can be neglected \cite{Soltanolkotabi}. Under these conditions, the total Hamiltonian of the system in a frame rotating at the laser frequency $\omega_L$ can be written as
\begin{equation}
\hat H = \hbar {\Delta _c}{\hat a^\dag }\hat a + \hbar {\omega _m}{\hat b^\dag }\hat b - \hbar {g_0}{\hat a^\dag }\hat a (\hat b + {{\hat b}^\dag }) +\!  i\hbar {E_L}({\hat a^\dag } - \hat a)\!,
\label{H1}
\end{equation}
where $\hat a$ ($\hat a^\dag$) and $\hat b$ ($\hat b^\dag$) denote, respectively, the annihilation (creation) operators of the single-mode cavity field and the mechanical motion of the MO, the parameter $\Delta_c=\omega_c-\omega_L$ is the detuning of the cavity mode from laser frequency, and $g_0=x_{\rm zpf} d\omega_c(q) / dq$ with $x_{\rm zpf}=\sqrt {\hbar /(2m{\omega _m})} $ being the zero-point position fluctuation, is the single-photon optomechanical coupling strength. The first and the second terms in Eq. (\ref{H1}) denote the cavity mode energy and the mechanical mode energy, respectively. The third term describes the optomechanical coupling of the MO to the cavity mode via radiation pressure. It is worthwhile to mention that the optomechanical coupling has been widely employed in a large variety of applications including detection and interferometry of gravitational waves \cite{LIGO}, displacement and force sensing \cite{xsensing1,CQNCPRL,CQNCPRX,CQNCmeystre,CQNCmaximilian,aliNJP}, ground state cooling of the MO \cite{ground state cooling,Sideband cooling,Laser cooling} and generation of nonclassical states of the mechanical and optical modes \cite{Borkje,Hammerer}. The last term in the Hamiltonian of Eq. (\ref{H1}) describes the driving of the intracavity mode with the input laser amplitude $E_L=\sqrt{\kappa P_L/\hbar \omega_L}$. 

The dynamics of the system is fully characterized by the fluctuation-dissipation processes affecting both the optical and the mechanical modes. We describe the effect of the fluctuations of the vacuum radiation input and the Brownian noise associated with the coupling of the MO to its thermal environment within the input-output formalism of quantum optics. For the given Hamiltonian (\ref{H1}) this results in the nonlinear quantum Langevin equations 
\begin{subequations} \label{langevin1}
	\begin{eqnarray}
	&& \dot {\hat a} =  - i{\Delta _c}\hat a + i{g_0}\hat a (\hat b + {{\hat b}^\dag }) + {E_L} - \frac{\kappa }{2}\hat a + \sqrt \kappa  {{\hat a}_{in}}, \\
	&& \dot {\hat b }=  - i{\omega _m}\hat b + i{g_0}{{\hat a}^\dag }\hat a  - \frac{{{\gamma _m}}}{2}\hat b + \sqrt {{\gamma _m}} {{\hat b}_{in}},
	\end{eqnarray}
\end{subequations} 
in which the cavity-field quantum vacuum fluctuation $\hat a_{in}$ and the motional quantum fluctuation $\hat b_{in}$ satisfy the commutation relations $\left[ {{{\hat a}_{in}}(t),\hat a_{in}^\dag ( t')} \right] =\left[ {{{\hat b}_{in}}(t),\hat b_{in}^\dag ( t')} \right]= \delta (t - t')$. In the limit of high mechanical quality factor ${Q_m} = {\omega _m}/{\gamma _m} \gg 1$ and when $\hbar {\omega _m} \ll {k_B}T$ where $k_B$ is the Boltzmann constant and $T$ is the temperature of the mechanical bath, $\hat b_{in}$ satisfies the nonvanishing Markovian correlation functions \cite{Tombesi} $\langle {\hat b_{in}^\dag (t){{\hat b}_{in}}(t')}\rangle  = {{\bar n}_m}\delta (t - t')$, $\langle {{{\hat b}_{in}}(t)\hat b_{in}^\dag (t')}\rangle  = ({{\bar n}_m} + 1)\delta (t - t')$ where ${{\bar n}_m} = {(\exp (\hbar {\omega _m}/{k_B}T) - 1)^{ - 1}}$ is the mean thermal excitation number of the MO. Furthermore, the nonvanishing correlation function of the input vacuum noise is given by \cite{Gardiner} $\langle {{{\hat a}_{in}}(t)\hat a_{in}^\dag (t')}\rangle  = \delta (t - t')$.

If the cavity is intensely driven so that the intracavity field is strong which is realized for high-finesse cavities and enough driving power, the quantum Langevin equations \ref{langevin1}(a) and \ref{langevin1}(b) can be solved analytically by adopting a linearization scheme in which the operators are expressed as the sum of their classical mean values and small fluctuations, i.e., $\hat a = \alpha + \delta \hat a$ and  $\hat b = \beta + \delta \hat b$ with $\langle \delta {{{\hat{r}}}^{\dag }}\delta \hat{r} \rangle / \langle {{{\hat{r}}}^{\dag }}\hat{r}\rangle \ll 1$ $(r=a, b)$. The classical amplitudes $\alpha  = \langle \hat a \rangle$ and $\beta  = \langle \hat b \rangle$ are governed by equations $\dot \alpha  =  - ({\kappa  \mathord{\left/{\vphantom {\kappa  2}} \right.\kern-\nulldelimiterspace} 2} + i{\bar \Delta _c})\alpha  + {E_L}$ and $\dot \beta  =  - ({{{\gamma _m}} \mathord{\left/ {\vphantom {{{\gamma _m}} 2}} \right. \kern-\nulldelimiterspace} 2} + i{\omega _m})\beta  + i{g_0}|\alpha {|^2}$ where ${\bar \Delta _c} = {\Delta _c} -2 {g_0} {\rm Re} \beta$. The dynamics of the quantum fluctuations can be described by the linearized quantum Langevin equations 
\begin{subequations} \label{fluctuation1}
	\begin{eqnarray}
	&& \delta {\dot {\hat a}} \! = \!  - (i{{\Delta }_c} + \frac{\kappa }{2})\delta {\hat a} + ig(\delta \hat b + \delta{{\hat b}^\dag })\! +\! \sqrt \kappa  {{\hat a}_{in}},\\
	&&  \delta \dot {\hat b} \! = \! - (i{\omega _m} \! + \! \frac{{{\gamma _m}}}{2}) \delta \hat b + i g(\delta \hat a + \delta{{\hat a}^\dag })\!  +\! \sqrt {{\gamma _m}} {{\hat b}_{in}},
	\end{eqnarray}
\end{subequations}
where $g = {g_0}{\alpha _{ss}} $ is the coherent intracavity-field-enhanced optomechanical coupling strength with $\alpha_{ss} = {E_L}/(\kappa /2 + i{{\bar \Delta }_c})$ being the steady-state value of $\alpha$ which is always possible to take as a real number by an
appropriate redefinition of phases. Assuming the driving field is sufficiently far-detuned away from the cavity resonance, i.e., $\left| {{\Delta _c}} \right| \gg \kappa ,{\omega _m},{g_0}$, one can approximate ${\bar \Delta _c} \approx {\Delta _c}$ for small values of the MO displacement which means that $\Delta _c \gg {g_0} Re \beta $. Furthermore, when the cavity decay rate $\kappa$ is much larger than the decay rate of the MO and the coupling strength ($\kappa \gg \gamma_m$, $g_0$),  the cavity field may be adiabatically eliminated on time scales greater than $\kappa^{-1}$. By integrating from both sides of Eq. \ref{fluctuation1}(a), neglecting the fast oscillating terms due to the large detuning, and inserting the resulting expression for $\delta \hat a(t)$ into Eq. \ref{fluctuation1}(b) one arrives at an effective equation of motion for the operator $\delta \hat b(t)$
\begin{eqnarray} \label{b_eff}
&& \delta \dot {\hat b}(t) \! =\!  - i (\omega _m \! + \frac{\gamma_m}{2}) \delta \hat b + 2i \dfrac{g^2}{\Delta_c}(\delta \hat b \! + \delta \hat b^\dag)  \! + \! {\hat F_b},
\end{eqnarray}
where 
\begin{eqnarray}
&& {\hat F_b} = ig\sqrt \kappa  \left( {{{\hat F}_a} + \hat F_a^{\rm{\dag }}} \right) + \sqrt {{\gamma _m}} {\hat b_{in}},
\label{Fb}
\end{eqnarray}
is the generalized non-Markovian noise operator with ${\hat F_a}(t) = {e^{ - \left( {i{\Delta _c} + \kappa /2} \right)t}}\int_0^t {dt'{{\hat a}_{in}}(t'){\rm{ }}{e^{\left( {i{\Delta _c} + {\kappa  \mathord{\left/
						{\vphantom {\kappa  2}} \right.
						\kern-\nulldelimiterspace} 2}} \right)t'}}}$. It should be noted that the first term in Eq. (\ref{Fb}) is a colored (non-Markovian) noise induced by the adiabatic elimination of the cavity mode, and the second term is a white
(Markovian) noise. Under the Markov approximation and for times longer than $\kappa^{-1}$ the noise operator $\hat F_a$ can be approximately written as $ {{\hat F}_a} \simeq 2{{\hat a}_{in}}(1 - {e^{ - (i{\Delta _c} + \kappa /2)t}})/(\kappa  + 2i{\Delta _c}) \simeq 2{{\hat a}_{in}}/(\kappa  + 2i{\Delta _c})$ which in the dispersive regime ($\vert \Delta_c \vert \gg \kappa, \omega_m,g_0$) becomes ${{\hat F}_a} \simeq (1+i \kappa/2\Delta_c) /(i{\Delta _c}){{\hat a}_{in}}$. In this manner Eq. (\ref{Fb}) can be simplified as
\begin{equation}
{{\hat F}_b} = \sqrt {{\gamma _m}} \left( {{{\hat b}_{in}} + \xi ( {{\hat a}_{in}}\delta - \hat a_{in}^\dag \delta ^* )} \right),
\label{Fbmarkov}
\end{equation}
where we have defined $\xi  = (g/{\Delta _c})\sqrt {\kappa /{\gamma _m}}$ and $\delta=1+i \kappa/2\Delta_c=1+i(\sqrt{\kappa \gamma_m }/2g) \xi$. As is clear, in the adiabatic limit ($\kappa \gg \gamma_m$), depending on the value of $g/\vert \Delta_c \vert$, two different cases can be identified: $\vert \xi \vert \gg1$ and $ \vert \xi \vert \ll 1$. In the first case which corresponds to $g/\vert \Delta _c \vert \gg \sqrt {\kappa /{\gamma _m}}  \gg 1$, one has ${{\hat F}_b} \approx  \sqrt {{\gamma _m}} \xi ({{\hat a}_{in}}\delta - \hat a_{in}^\dag \delta^*)$ whereas in the second case, corresponding to $g/\vert \Delta _c \vert \ll 1 \ll \sqrt {\kappa /{\gamma _m}}$, one gets ${{\hat F}_b} \approx \sqrt {{\gamma _m}} {{\hat b}_{in}}$. Both cases are generally achievable by controlling the effective linearized optomechanical coupling strength $g=g_0 \alpha_{ss}$ through input laser power $P_L$(of course, due to the dispersive approximation we have always $g_0 \ll \vert \Delta_c \vert$). Thus, the parameter $\xi$ can be identified as an effective control parameter by which the noise induced on the MO can be controlled. However, the experimental realization of the system under consideration shows $\vert \xi \vert \ll 1$ (e.g., for the experimental values given in Ref. \cite{Thomson} we obtain $\vert \xi \vert \sim 10^{-3}$). Therefore, we keep terms up to second order in $\xi$ since, as we will show later, the leading terms dependent on this parameter appearing in the relevant physical quantities such as the generated Casimir phonon number and the squeezing parameter are of order $\xi^2$. Within this approximation, the noise operator $\hat F_b$ satisfies the commutation relation $\left[ {{{\hat F}_b(t)},\hat F_b^\dag (t')} \right] = {\gamma _m}\delta (t - t')$ and obeys the correlation functions  $\langle {{{\hat F}_b(t)}{{\hat F}_b(t')}}\rangle  =  - {\gamma _m}{\xi ^2}\delta (t - t')$, $\langle {{{\hat F}_b(t)}\hat F_b^\dag(t') }\rangle  = {\gamma _m}\left( {(1 + {\bar n_m}) + {\xi ^2}} \right)\delta (t - t')$ and $\langle \hat F_b^\dag(t) \hat F_b(t') \rangle={\gamma _m}\left( {\bar n_m+ {\xi ^2}} \right)\delta (t - t') $. 

It is clear from the equation of motion (\ref{b_eff}) that in the dispersive regime and under adiabatic elimination of the cavity field the dynamics of quantum fluctuations of the MO in the Schr\"odinger picture is governed by the effective Hamiltonian 
\begin{equation}
{{\hat H}_{eff}} = \hbar {\Omega _m}\delta{{\hat b}^\dag }\delta \hat b - \hbar {\chi _0}(\delta{{\hat b}^2} + \delta{{\hat b}^{\dag 2}}),
\label{H_effective}
\end{equation}
in which $\chi_0={g^2}/{\Delta _c}$ and ${\Omega _m} = {\omega _m} - 2{\chi _0}$ is the shifted frequency of the MO due to the radiation pressure, known as the optical spring effect \cite{Genes}. In the rotating frame at frequency $\Omega_m$ and within the rotating wave approximation (RWA) the effective Hamiltonian (\ref{H_effective}) reduces to zero Hamiltonian of the form
\begin{equation}
\mathcal{H}_{eff,RWA}^{(\rm I)}= \hbar {\chi _0}(\delta {{\hat b}^2} e^{-2i\Omega_m t} + \delta {{\hat b}^{\dag 2}} e^{+2i\Omega_m t}) \approx 0 .
\label{HR}
\end{equation}

We now assume that the cavity is driven by a pump field with harmonically modulated power ${P_L}(t) = {P_L}(1 + \varepsilon \cos (\Omega t))$. Because of the adiabatic elimination of the optical field the cavity decay can be avoided and thus we consider the time scales greater than $\kappa^{-1}$. The modulation of the input power implies that we will now take ${\chi _0} \to \chi (t) = {\chi _0}(1 + \varepsilon \cos (\Omega t))$. Accordingly, the Hamiltonian (\ref{H_effective}) takes the form 
\begin{equation}
\hat {\tilde H}_{eff} = \hbar {\Omega _m}(t)\delta{{\hat b}^\dag }\delta\hat b - \hbar \chi (t)(\delta{{\hat b}^2} +\delta {{\hat b}^{\dag 2}}),
\label{H rewrite modulate}
\end{equation}
where ${\Omega _m}(t) ={\Omega _m} - 2{\chi _\varepsilon }\cos (\Omega t)$ with $\chi_\varepsilon=\varepsilon \chi_0$. 
Now, by choosing the condition such that $\Omega  =2 \Omega_m= 2({\omega _m} - 2\chi_0)$ and applying the RWA to neglect the fast oscillating terms $\delta{{\hat b}^2}{e^{ - 2i{\Omega _m}t}}$, $\delta{{\hat b}^{\dag 2}}{e^{2i{\Omega _m}t}}$, and ${\chi _\varepsilon }\cos (\Omega t)\delta{{\hat b}^\dag }\delta \hat b$ which can be satisfied when one looks at the system on a time scale larger than mechanical frequency, $t \gg \omega_m^{-1}$, one can obtain the effective Hamiltonian in the frame rotating at the shifted frequency of the MO, $\Omega_m$, as follows
\begin{equation}
\hat {\tilde H}_{eff,RWA}^{(I)}=  - \hbar \frac{{{\chi _\varepsilon }}}{2}(\delta{{\hat b}^2} + \delta{{\hat b}^{\dag 2}}),
\label{H_eff_squeezing_final_RWA}
\end{equation}
or in the Schr{\"o}dinger picture
\begin{equation} \label{H_eff_squeezing_final_RWA_Schrödinger}
\hat {\tilde H}_{eff,RWA} \! =\! \hbar {\Omega _m}\delta{{\hat b}^\dag }\delta\hat b - \hbar \frac{{{\chi _\varepsilon }}}{2}(\delta{{\hat b}^2}{e^{2i\Omega_m t}}\! + \! \delta {{\hat b}^{\dag 2}}{e^{ - 2i\Omega_m t}}).
\end{equation}

The Hamiltonian of Eq. (\ref{H_eff_squeezing_final_RWA}) describes a phonon analog of the degenerate optical parametric amplification (DPA) where the vibrational fluctuation of the MO plays the role of the signal mode in the DPA. Besides, the parameter $\chi_\varepsilon$, which is controllable through the optomechanical coupling strength $g$, detuning $\Delta_c$, and amplitude of modulation $\varepsilon$, acts as the nonolinear gain parameter. Accordingly, it is expectable that in the far-detuned regime of cavity pumping with an amplitude-modulated drive the amplification of quantum fluctuations of the MO, i.e., the generation of Casimir phonons, will be possible over time scales longer than $\kappa^{-1}$.
It is emphasized that our proposed scheme for generation of phonons is different from the well-known process of nondegenerate (spontaneous) phonon emission \cite{Aspelmeyer} where energy is transferred between the driving laser and the joint excitations of the optical and the mechanical mode, resulting in the parametric amplification of the mechanical oscillator (i.e., the parametric oscillatory instability \cite{LIGO}.
It should be noted that although Eq. (\ref{b_eff}) requires only $\vert \Delta_c \vert \gg \kappa$, Eqs. (\ref{H_eff_squeezing_final_RWA}) and (\ref{H_eff_squeezing_final_RWA_Schrödinger}) require that $\vert \Delta_c \vert \gg \kappa \ge \omega_m$ (unresolved sideband and far-detuned regimes). Moreover, working in the far-detuned regime together with the single-cavity mode approximation one needs to consider quite short cavities such that the separation between the optical modes, i.e., free spectral range ($\rm FSR$), is larger than the cavity detuning, i.e., $ {\rm FSR} \gg \vert \Delta_c \vert  $ where ${\rm FSR}=\kappa \mathcal{F}$ and the coefficient of finesse $ \mathcal{F}$ in the Fabry-P\'{e}rot cavity is defined as $\mathcal{F}=\pi c/(2\kappa L)$. In the following, we determine the  dynamics of the quadrature fluctuations of the MO by which we will calculate the mean number of generated Casimir phonons and will investigate their quantum statistical properties in the subsequent sections.

Generally, the Langevin equation for the fluctuation operator $\delta \hat b$ in the interaction picture is given by 
\begin{equation}
\delta \dot {\hat b} = \! - i{\Omega _m}(t)\delta \hat b - \frac{{{\gamma _m}}}{2} \delta\hat b + 2i\chi (t)\delta {{\hat b}^\dag } \! + \! {{\hat F}_b}(t),
\label{b general}
\end{equation}
from which one can obtain the following equations of motion for the canonical mechanical quadrature fluctuation
operators $\delta \hat X= (\delta \hat b + \delta \hat b^\dag)/2$ and $\delta\hat P= i(\delta\hat b^\dag -\delta \hat b)/2$,
\begin{subequations} \label{XP general}
	\begin{eqnarray}
	&& \delta \dot {\hat X} = \left( {2\chi (t) + {\Omega _m}(t)} \right)\delta \hat P - \frac{{{\gamma _m}}}{2}\delta\hat X + {{\hat X}_{in}}, \\
	&& \delta\dot {\hat P }= \left( {2\chi (t) - {\Omega _m}(t)} \right)\delta \hat X - \frac{{{\gamma _m}}}{2}\delta\hat P + {{\hat P}_{in}},
	\end{eqnarray}
\end{subequations}
where ${{\hat X}_{in}} = ({{\hat F}_b} + \hat F_b^\dag )/2$ and ${\hat P}_{in} = i(\hat F_b^\dag  - {{\hat F}_b})/2$. In the interaction picture and under RWA the above equations of motion take the form
\begin{subequations} \label{X and P equation}
	\begin{eqnarray}
	&& \delta \dot {\hat X} = {\chi _\varepsilon }\delta \hat P - \frac{{{\gamma _m}}}{2}\delta \hat X + {{\hat X}_{in}}, \\
	&& \delta \dot {\hat P} = {\chi _\varepsilon }\delta \hat X - \frac{{{\gamma _m}}}{2}\delta \hat P + {{\hat P}_{in}} ,
	\end{eqnarray}
\end{subequations} 
which can be expressed in the compact matrix form
\begin{equation}
\dot {\hat u}(t) = A\hat u(t) + \hat N(t),
\label{udot}
\end{equation}
where the vector fluctuation and the corresponding vector noise are $\hat u(t) = {(\delta\hat X(t),\delta\hat P(t))^T}$ and $\hat N(t) = {({{\hat X}_{in}},{{\hat P}_{in}})^T}$, respectively. Furthermore, the time-independent drift matrix $A$ is given by 
\begin{equation}
A = \left( {\begin{array}{*{20}{c}}
	{ - \frac{{{\gamma _m}}}{2}}&{{\chi _\varepsilon }}\\
	{{\chi _\varepsilon }}&{ - \frac{{{\gamma _m}}}{2}}
	\end{array}} \right).
\label{A matrix}
\end{equation}
Stability of the dynamics encompassed by Eq. (\ref{udot}) can be obtained from the Routh-Hurwitz criterion \cite{Ruth-Hurwitz} which requires that ${\gamma _m} > 2\left| {{\chi _\varepsilon }} \right|$. It should be mentioned that since detuning can be negative (blue detuning) or positive (red detuning), $\chi_\varepsilon$ can be negative or positive. The formal solution of Eq. (\ref{udot}) is given by 
\begin{equation}
\hat {u}(t) = {e^{A(t - {t_0})}}\hat {u}({t_0}) + \int_{{t_0}}^t {dt'{e^{A(t - t')}}\hat N(t')}.
\label{u}
\end{equation}
By solving the eigenvalue problem for the matrix $\rm A$ we find the eigenvalues ${\lambda _{1,2}} =  - {\gamma _m}(1 \pm 2s_\varepsilon)/2$ where $s_\varepsilon = {\chi _\varepsilon }/{\gamma _m}$, and the eigenvectors $\left| {{\lambda _{1,2}}} \right\rangle  = 1/\sqrt 2 {(1, \mp 1)^T}$. It should be mentioned that the stability condition in terms of $s_\varepsilon$ is given by $\left| {{s _\varepsilon }} \right| <1/2$. In our formulation the red and blue detuning cases are, respectively, associated with positive and negative values of $s_\varepsilon$. It is evident that the replacements $s_\varepsilon \to -s_\varepsilon$ and $\lambda_1  \leftrightarrow  \lambda_2$ are equivalent to moving from the red-detuned ($\Delta_c>0$) to the blue-detuned ($\Delta_c<0$) regime.
By using a unitary transformation with the time-independent unitary matrix $R = \left( {\left| {{\lambda _1}} \right\rangle ,\left| {{\lambda _2}} \right\rangle } \right)$, Eq. (\ref{udot}) is transformed into 
\begin{equation}
\dot {\hat {\tilde u}} = {A_d}\hat {\tilde u} + \hat {\tilde N} ,
\label{u tilde dot}
\end{equation}
where $\hat {\tilde u} = {R^\dag }\hat u = 1/\sqrt 2 {(\delta\hat X - \delta\hat P,\delta\hat X + \delta\hat P)^T}$, $\hat {\tilde N} = {R^\dag }\hat N = 1/\sqrt 2 {({{\hat X}_{in}} - {{\hat P}_{in}},{{\hat X}_{in}} + {{\hat P}_{in}})^T}$, and ${A_d} = {R^\dag }AR = {\rm Diag}({\lambda_1},{\lambda_2})$. The solution of the transformed Eq. (\ref{u tilde dot}) is given by 
\begin{eqnarray}
&& \! \! \! \! \! \! \! \! \! \! \! \! \! \!\! {{\hat {\tilde u}}_i}(t) \! \! =\!  {e^{{\lambda _i}(t - {t_0})}} {{\hat {\tilde u}}_i}({t_0}) \!   + \!  \!  \int_{{t_0}}^t \! dt'  {{e^{{\! \lambda _i} \! (t - t') \!}}  {{\hat {\tilde N}}_i} \! (t')} ~ ~ (i=\! \! 1,2) .
\label{ui tilde}
\end{eqnarray}
In particular, the transformed symmetric covariance matrix $\tilde V$ with entries given by ${\tilde V_{ij}}(t): = \langle {{\hat {\tilde u}_i}(t){\hat {\tilde u}_j}(t) + {\hat {\tilde u}_j}(t){\hat {\tilde u}_i}(t)}\rangle /2$ fully characterizes the quantum dynamics of the quadrature fluctuations of the MO. From Eq. (\ref{ui tilde}) the equation of motion of $\tilde V$ is obtained as 
\begin{equation}
\frac{{d \tilde V(t)}}{{dt}} = A_d \tilde V(t) + \tilde V(t){A_d^{T}}  + \tilde D,
\label{Vdot}
\end{equation}
where $\tilde D$ is the diffusion matrix, the matrix of noise correlation, defined as 
\begin{equation}
{{\tilde D}_{ij}}\delta (t - t'){\kern 1pt}  = \frac{1}{2}\langle {{{\hat {\tilde N}}_i}(t){{\hat {\tilde N}}_j}(t') + {{\hat {\tilde N}}_j}(t'){{\hat {\tilde N}}_i}(t)}\rangle ,
\label{Dtilde}
\end{equation}
with ${{\tilde D}_{11}} ={{\tilde D}_{22}} = \gamma_m /4 \mathcal{C}(1 + 2{\bar n_m})$ and ${{\tilde D}_{12}} = {{\tilde D}_{21}} = (1-\mathcal{C})/4$ where $\mathcal{C}= 1 + 2{\xi ^2}/(1 + 2{{\bar n}_m})$.

\subsection{Mean number of generated Casimir phonons}

Having determined the temporal evolution of the mechanical quadrature fluctuations, we now proceed to calculate the mean number of generated Casimir phonons. We assume that the precooled MO is initially in the thermal state, i.e., ${{\tilde V}_{ii}}(0) = 0.25(1 + 2{{\bar n}_m})$, then 
\begin{eqnarray}
&& \! \! \! \! {{\tilde V}_{1,2}}(t) \! = \! \frac{{1\! +\! 2{{\bar n}_m}}}{4} \! \left[ {\frac{\mathcal{C}}{{1 \pm 2{s_\varepsilon }}} \! + {e^{2{\lambda _{1,2}}t}}\! (\frac{{1\! - \! \mathcal{C} \! \pm 2{s_\varepsilon }}}{{1 \pm 2{s_\varepsilon }}})} \right] \! ,
\label{V tilde solve}
\end{eqnarray}
where ${{\tilde V}_i} \equiv {{\tilde V}_{ii}}$. The mean number of generated phonons at time $t$ can be extracted from the covariance matrix, $n(t) = \langle \delta \hat {\tilde b}^\dag \delta \hat {\tilde b} \rangle  = \tilde V_1 + \tilde V_2 - 1/2$, which reads as
\begin{eqnarray}
&& \!  \! \! \! \! \! \! \! \! \! \! \! \! \! \! \! \! n(t) \!=\! n_{ss} \! + \! \frac{\! 1 \! + \! 2{{\bar n}_m} \! }{4} \! \!  \left[ \! \frac{{1 \! - \! {\cal C} \! + \! 2{s_\varepsilon }}}{{1 + 2{s_\varepsilon }}}{e^{2{\lambda _1}t}} \!  + \!  \frac{{1 \! - \! {\cal C} \! - \! 2{s_\varepsilon }}}{{1 - 2{s_\varepsilon }}}{e^{2{\lambda _2}t}} \! \right] \! ,
\label{nb}
\end{eqnarray}
where the steady-state mean value is given by ${n_{ss}} = \left( {2({\cal C} - 1) + 4{\cal C}{{\bar n}_m} + 8s_\varepsilon ^2} \right)/\left( {4(1 - 4s_\varepsilon ^2)} \right)$. Equation (\ref{nb}) shows that, as expected, the initial mean phonon number is equal to the bath phonon number, i.e, $n(0)=\bar n_m$. Moreover, in the absence of time modulation of the driving laser Eq. (\ref{nb}) reduces to $n(t) \vert_{\varepsilon=0}= \bar n_m +\xi^2 (1-e^{-\gamma_m t}) \cong \bar n_m =n(0)$ (the parameter $\xi$ is vanishingly small). 

Since we would like to study the generated Casimir phonons due to amplification of the mechanical vacuum fluctuations, we separate the contributions of the generated Casimir and the thermal phonons in Eq. (\ref{nb}) as $ n(t) = n_{Casimir}(t)+ n_{th}(t)$, where
\begin{eqnarray}
&& \! \! \!  \! \! \!  \!  \! \! \!  \! \! \!  n_{Casimir}(t) \!= \! \frac{2s_\varepsilon^2 \! + \! \xi^2}{1-4s_\varepsilon^2} \! + \! \frac{1}{2} \! \! \left[ \! \frac{\! s_\varepsilon \! - \! \xi^2}{1 \!+ \! 2s_\varepsilon} e^{2\lambda_1 t} \! - \! \frac{\! s_\varepsilon \! + \! \xi^2}{1\!-\!2s_\varepsilon} e^{2\lambda_2 t} \! \right] \!\!, \label{ncas} \\
&& \! \! \! \! \! \! \!  \! \! \!  \! \! \!   n_{th}(t)\!=\! \bar n_m \left[ \frac{1}{1-4s_\varepsilon^2} + s_\varepsilon \left( \frac{e^{2\lambda_1 t}}{1+2s_\varepsilon}  \! - \frac{e^{2\lambda_2 t}}{1-2s_\varepsilon} \right)   \right] \!.
\label{nthermal}
\end{eqnarray}
\begin{figure}
	\includegraphics[width=8.5cm,height=2.7cm]{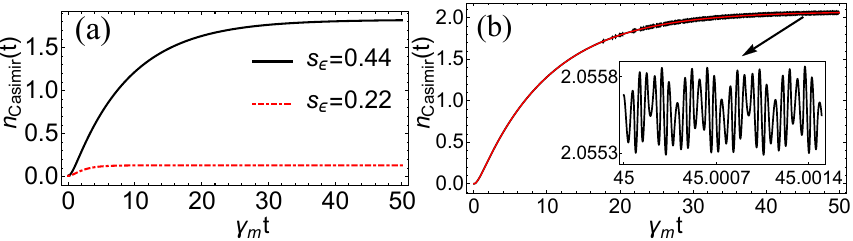}
	\caption{(Color online) (a) Mean number of generated Casimir phonons obtained analytically within the RWA as a function of $\gamma_m t$ (Eq. (\ref{ncas})) for two positive values of $s_\varepsilon$ (red-detunned regime). (b) Comparison of the temporal behaviour of $n_{Casimir}(t)$ as predicted by RWA (red line) and the result given by the numerical calculation beyond the RWA (black line) through the solution of Eq. (\ref{b general}) including the CRTs for $s_\varepsilon=0.44$. The inset shows the tiny and fast oscillations resulting from the CRTs.} 
	\label{Fig2}
\end{figure}
To examine the effect of the external pump modulation on the mean number of generated Casimir phonons, in Fig. \ref{Fig2}(a) we have plotted $n_{Casimir}(t)$ obtained analytically within the RWA, versus the scaled time $\gamma_m t$ for two positive values of $s_\varepsilon$ (red-detuned regime). As can be seen, with increasing $s_\varepsilon$ toward its maximum value ($<0.5$) the rate of Casimir phonon generation increases. This shows that time-modulated pumping of the MO is responsible for the generation of Casimir phonons. In other words, energy can be transfered from coherent external driving to the mechanical vacuum fluctuations and leads to its parametric amplification.  The number of generated phonons can be controlled via the pump power $P_L$, detuning $\Delta_c$, and amplitude of pump modulation $\varepsilon$. In fig. \ref{Fig2}(b) we illustrate the effect of CRTs on the temporal behaviour of the mean number of generated phonons . From this figure, we find that taking into account the CRTs does not lead to any difference compared to the result predicted by the RWA, except for tiny and fast oscillations in the long-time limit. As can be seen from Eq. (\ref{nthermal}), $ n_{th}^{ss}/n_{Casimir}^{ss}=0.5 \bar n_m / s_\varepsilon^2 \le 2 \bar n_m $, which indicates that the steady-state mean number of generated thermal phonons can be greater than or equal to the mean number of generated Casimir phonons if $ \bar n_m \ge 2 s_\varepsilon^2   $. 

Here, we have used typical experimental values for the optomechanical parameters reported in \cite{Thomson} for a SiN membrane-in-the-middle optomechanical cavity with length $ L=6.7 $cm, mass $ m=4 \times 10^{-8} ${\rm g}, lowest frequency $ \omega_m/2\pi=134 $KHz, damping rate $ \gamma_m/2\pi= 0.12$Hz, cavity decay rate $ \kappa/2\pi=5 \times 10^5 $Hz and $ {\rm FSR} \simeq 2.8 \times 10^{10} $ Hz with optomechanical strength $ g_0/2\pi=50 $Hz. The reflectivity of the membrane and its size are $ \vert r_c \vert=0.42 $ and $ 1 $mm $\times 1 $mm $\times 50 $nm, respectively. The membrane can be initially cryogenically precooled to $ T=300 $mK. It should be pointed out that the ground-state precooling of the MO can be experimentally achieved \cite{Thomson} based on the quantum theory of cavity-assisted sideband cooling \cite{marquartcooling}: after cryogenically precooling of the membrane in the resolved sideband and weak coupling regimes \cite{Thomson,marquartcooling} by an auxiliary input laser power $ P_{in} \simeq 0.1$nW, it can be cooled down to its ground state ($ \bar n_m \le 0.1 $) by using an auxiliary cavity mode with $ \kappa_{\rm aux}/2\pi \simeq 8  $KHz. After ground-state cooling the membrane remains only in the weak coupling regime ($ \Gamma_{opt} \ll \kappa_{\rm aux} \sim \omega_m $). 

As is evident from Eqs. (\ref{nb}-\ref{nthermal}), $n_{Casimir}(t)$ ($ n(t) $) is invariant under the replacements $s_\varepsilon \to -s_\varepsilon$ and $\lambda_1 \leftrightarrow \lambda_2$, which indicates that the mean number of generated phonons is equal for both cases of red- and blue- detuned regimes. However, the phonons generated in the red-detuned regime have lower frequency than those generated in the blue-detuned regime such that $\Omega_{Casimir}^{\rm {red}}=\omega_m - 2  \vert \chi_0 \vert$ and $\Omega_{casimir}^{\rm {blue}}=\omega_m + 2  \vert \chi_0 \vert$. This difference is expectable since on the red-detuning side the optomechanical coupling generates a net energy transfer from the MO into the cavity while on the blue-detuning side, the driving laser gives rise to an energy transfer into the MO. 
 
\begin{figure*}
	\includegraphics[width=18.5cm,height=3.6cm]{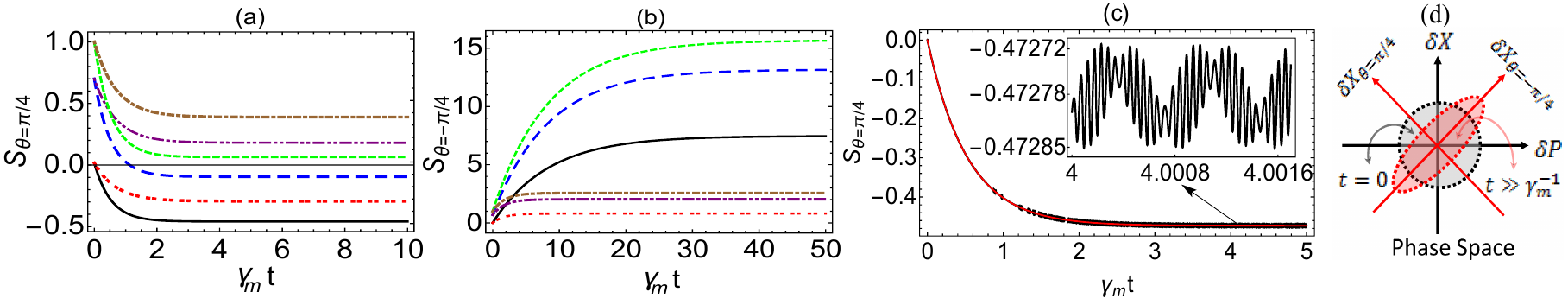}
	\caption{( Color online) (a) Squeezing parameter $S_1=S_{\theta=\pi/4}$ and (b) squeezing parameter $S_2=S_{\theta=-\pi/4}$ obtained analytically within the RWA versus the scaled time $\gamma_m t$ for two positive values of $s_\varepsilon$ (red-detuned regime). The black solid line is for $ s_\varepsilon=0.44$ and $ \bar n_m \approx 0.01$, the red-double-dotted line is for $ s_\varepsilon=0.22$ and $ \bar n_m \approx 0.01$, the blue long-dashed line is for $ s_\varepsilon=0.44$ and $ \bar n_m=0.35$, the purple double dot-dashed line is for $ s_\varepsilon=0.22$ and $ \bar n_m \approx 0.35$, the green dashed line is for $ s_\varepsilon=0.44$ and $ \bar n_m \approx 0.5$ and the brown dot-dashed line is for $ s_\varepsilon=0.22$ and $ \bar n_m \approx 0.5$. (c) Comparison of the temporal behaviour of the squeezing parameter $S_{\theta=\pi/4}$ as predicted by the RWA (red line) and the result given by the numerical calculation beyond the RWA (black line) through the solution of Eq. (\ref{b general}) including the CRTs for $s_\varepsilon=0.44$ and $ \bar n_m=0.01 $. The inset shows the tiny and fast oscillations due to the CRTs. (d) Schematic phase-space picture of mechanical squeezing. A MO in thermal equilibrium with equal variance in conjugate quadratures (black dotted circle) is dispersively pumped by a coherent modulated driving field at frequency $\omega_L$ and power $P_L$ that drives the MO into a state with squeezed variance of the quadrature $\delta \hat X_{\theta=\pi/4}$ (red dotted ellipse).} 
	\label{Fig3}
\end{figure*}
\section{ QUANTUM STATISTICAL PROPERTIES OF THE GENERATED CASIMIR PHONONS} \label{sec3}

\subsection{Phonon quadrature squeezing}
If we define the generalized quadrature fluctuation operators $\delta{{\hat X}_\theta } = (\delta\hat b{e^{i\theta }} + \delta{{\hat b}{^\dag} }{e^{ - i\theta }})/2$ and $\delta{{\hat P}_\theta } = i(\delta{{\hat b}^\dag }{e^{ - i\theta }} - \delta \hat b{e^{i\theta }})/2$ with $\delta \hat {\tilde b} \equiv \delta \hat b{e^{i\theta }}$, then ${{\hat {\tilde u}}_{1,2}} = \delta{{\hat X}_{\theta  = \pi /4, - \pi /4}}$. If either ${\rm Var} ( \delta \hat X_{\theta=\pi/4} )$ or ${\rm Var} (\delta \hat X_{\theta=-\pi/4})$, where ${\rm Var}(\delta \hat X_\theta)= \langle (\delta \hat X_\theta)^2 \rangle - \langle \delta \hat X_\theta \rangle ^ 2$, is less than $1/4$ the state of the MO exhibits quadrature squeezing. The degree of squeezing can be measured by the squeezing parameter $S_i$ ($i=1,2$) defined by $S_i=4{\rm Var}(\delta \hat X_\textit{i}) - 1=4\tilde V_\textit{i} -1$ where $i=1$ and $i=2$ correspond, respectively, to $\theta=\pi/4$ and $\theta=-\pi/4$. Then, the condition for squeezing in the quadrature component can be simply written as $S_i<0$. In the absence of the time modulation of the driving laser ($\varepsilon=0$), one has $s_\varepsilon=0$, $\lambda_1=\lambda_2=\gamma_m/2$ and consequently the squeezing parameter reads as $S_1 \vert_{\varepsilon=0} =S_2 \vert_{\varepsilon=0}=2\bar n_m + 2 \xi^2 (1-e^{- \gamma_m t}) > 0$, which shows that neither $S_1$ nor $S_2$ shows quadrature squeezing in the course of time evolution. Numerical results corresponding to the time evolution of the squeezing parameter $S_1$ and $S_2$ obtained analytically within the RWA for the case of time modulated pumping ($\varepsilon \neq 0$) and for two positive values of $s_\varepsilon$ (red-detuned regime) are, respectively, shown in Figs. \ref{Fig3}(a) and \ref{Fig3}(b) based on the experimentally feasible parameters given in Ref. \cite{Thomson}.
As can be seen, in this case one of the quadratures is always squeezed ($S_{\theta=\pi/4}<0$) if $ \bar n_m \ll 1 $. In the steady state, the squeezing parameter reaches $\left( {({\mathcal C} - 1) + 2{\mathcal C}{{\bar n}_m} \mp 2{s_\varepsilon }} \right)/(1 \pm 2{s_\varepsilon })$. As can be seen, by increasing $ \bar n_m $ the quadrature squeezing rapidly disappears. It is evident that if ${{\bar n}_m} \ll 1$, then $\%50$ noise reduction in the quadrature $\delta \hat X_{\theta=\pi/4}$ is achievable when $s_\varepsilon \to 0.5$. 
Based on the result obtained in Fig. \ref{Fig3}(a) one can control the quadrature squeezing of the generated Casimir phonons by the parameter $s_\varepsilon$ which itself is controllable through the pump power $P_L$, detuning $\Delta_c$, and amplitude of pump modulation $\varepsilon$, because ${s_\varepsilon } = 0.5\varepsilon ({P_L}/{P_0}) g_0^2/({\Delta _c} \gamma _m) $ where ${P_0} = \hbar {\omega _L}\Delta _c^2/\kappa $. In Fig. \ref{Fig3}(c) we compare the analytical solution based on RWA against the numerical result beyond the RWA, obtained from the solution of Eq. (\ref{b general}) including the CRTs. As is seen clearly, incorporation of the CRTs, apart from tiny and fast oscillations in the long-time behaviour of the mechanical squeezing, does not produce any difference compared to the result predicted by the RWA.
The phase-space description of the mechanical squeezing is shown in Fig. \ref{Fig3}(d). It is clear from Eq. (\ref{V tilde solve}) that if $s_\varepsilon \to - s_\varepsilon$ the mechanical squeezing is transferred from the  quadrature $\delta \hat X_{\theta=\pi/4}$ to the quadrature $\delta \hat X_{\theta=-\pi/4}$. In other words, the mechanical quadratures $\delta \hat X_{\theta=\pi/4}$ and $\delta \hat X_{\theta=-\pi/4}$ exhibit squeezing, respectively, in the red-detuned and blue-detuned regime.

The generated motional squeezing is vulnerable to the membrane environment temperature such that for temperatures higher than a critical value, $T_{\rm cr}$, no quadrature squeezing of the membrane occurs. To determine $T_{\rm cr}$, we rewrite the squeezing parameter for the red-detuned regime ($s_\varepsilon >0$) as ${S_i} = A_{ss}^ \pm  + {B_ \pm }{e^{2{\lambda _i}t}}$ ($i=1,2$ where $i=1$ corresponds to ``$+$" and $i=2$ corresponds to ``$-$") with $ A_{ss}^ \pm  = ({{\mathcal{C}} - 1 + 2{\mathcal{C}}{{\bar n}_m} - ( \pm 2{s_\varepsilon })}) / (1 \pm 2s_\varepsilon) $ and $ {B_ \pm } = (1 + 2 \bar n_m) (1 - \mathcal{C} \pm 2 s_\varepsilon ) / (1 \pm 2s_\varepsilon)$.
The squeezing criterion ($S_i <0$) requires that the conditions $A_{ss}^ \pm \leq 0 $ and $B_\pm \leq - A_{ss}^ \pm $ together with $\bar n_m \geq 0$ are satisfied simultaneously. It is straightforward to show that in the case of red-detuning ($0<s_\varepsilon<1/2$) these conditions can only be satisfied for $i=1$ resulting in a constraint on the thermal excitation number $\bar n_m$, i.e., $0 \leq{{\bar n}_m} \le (1/2)\left( {(1 - {\mathcal{C}} + 2{s_\varepsilon })/(3{\mathcal{C}} - 1 + 2{s_\varepsilon })} \right)$. It is obvious that for the case of blue detuning ($-1/2<s_\varepsilon<0$) the squeezing can only be occurred in the second quadrature ($i=2$). Therefore, considering both cases of red- and blue- detuned regime, the squeezing condition can generally be expressed as
\begin{eqnarray}
&& {{\bar n}_m} \le n_{\rm cr} = \frac{1}{2}\frac{{1 - {\cal C} + 2\left| {{s_\varepsilon }} \right|}}{{3{\cal C} - 1 + 2\left| {{s_\varepsilon }} \right|}},
\label{n_m}
\end{eqnarray}
in which $((\mathcal{C}-1)/2)\leq \left| {{s_\varepsilon }} \right| <1/2$. By using the relation  $\mathcal{C}=1+2\xi^2/(1+2\bar n_m)$ and making some algebraic simplification, the critical thermal excitation number $n_{\rm cr}$ can be written as $ n_{\rm cr}= (\sqrt{(1 + 3\xi^2)^2 + 4(\xi^2 +  \vert s_\varepsilon \vert )(1 +\vert s_\varepsilon \vert )}  - ( 1 + 3\xi^2  )) / (4 (1 + \vert s_\varepsilon \vert ) )$. 
In the limiting case $\vert s_\varepsilon \vert \to 1/2$ and $\mathcal{C} \to 1$ one finds $n_{\rm cr}=1/6 \approx 0.166$. Now by considering  ${{\bar n}_m} = {(\exp (\hbar {\omega _m}/{k_B}T) - 1)^{ - 1}}$, we can express the squeezing condition in terms of temperature as $T \leq T_{\rm {cr}}$ where the critical temperature is given by $T_{\rm {cr}}=\hbar \omega_m /(k_B {\rm ln}(1+ n_{\rm cr}^{-1}))$.

Based on Eq. (\ref{n_m}) in order that the generated Casimir phonons exhibit quadrature squeezing not only at steady state but also during transient state, the MO should be precooled down to its ground state that was explained in the previous section. Moreover, numerical calculations shows that if $ n_m \gtrsim 0.45$ no squeezing occurs even in the steady state.

Here, it is worth to mention that the generated mechanical squeezing in the present scheme can be detected based on an experimentally feasible method proposed in Ref. \cite{Vitali.statistic measurement}. The mechanical fluctuation quadratures $ \delta \hat X $ and $ \delta \hat P $ can be measured by homodyning the output field of another ancilla cavity mode with an appropriate phase, and driven by a much weaker pump laser so that its back-action on the mechanical mode can be ignored. Therefore, the ancilla mode adiabatically follows the membrane dynamics. Measuring the quadrature fluctuation of the output field of the ancilla cavity mode can thus be used to extract information about the mechanical quadrature squeezing.

\subsection{phonon counting statistics}
We now proceed to study the counting statistics of the generated Casimir phonons, including autocorrelation function and the Mandel parameter. In particular, we will show how one can manipulate the phonon statistics through controlling the external modulated pumping and the optomechanical parameters. 

The autocorrelation function is proportional to the probability of detecting one phonon at time $t+\tau$, given that another phonon was detected at earlier time $t$. It is defined by 
\begin{eqnarray} \label{g2}
&& \! \!\! \! \! \!\! \!  \! \! \! {g^{(2)}}(\tau )\!=\! \! \langle {\hat \delta {\tilde b}^{\dag} \! (t) \delta{\hat {\tilde b}^{\dag}}(t \! + \! \tau )\delta \hat {\tilde b}(t \! + \! \tau )\delta \hat {\tilde b}(t)} \rangle \! / \! \langle \hat n(t) \rangle^2 \! .
\end{eqnarray}
When ${g^{(2)}}(\tau )/{g^{(2)}}(0) < 1$ the phonons tend to distribute themselves preferentially in bunches rather than at random (phonon bunching). On the other hand, if ${g^{(2)}}(\tau )/{g^{(2)}}(0) > 1$, fewer phonons pairs are detected close together than further apart (phonon antibunching). Using the Gaussian properties of the noise forces \cite{Walls} in the steady-state, $t \gg \gamma _m^{ - 1}$, the autocorrelation function in Eq. (\ref{g2}) can be written as 
\begin{equation}
{g^{(2)}}\! (\tau ) \!  =\! 1 \! +\!  \frac{{{\vert {\langle \delta{{{\hat {\tilde b}}^\dag }\!\!(t)\delta{{\hat {\tilde b}}^\dag }\! \!(t \! + \! \tau )}\rangle }\vert^2}}}{{{\langle \delta{{{\hat {\tilde b}}^\dag }\!\!(t)\delta\hat {\tilde b}(t)}\rangle ^2}}} + \frac{{{\vert {\langle \delta{{{\hat {\tilde b}}^\dag }\! \! (t)\delta \hat {\tilde b}(t \! + \! \tau )}\rangle }\vert^2}}}{{{\langle \delta{{{\hat {\tilde b}}^\dag } \! \! (t)\delta\hat {\tilde b}(t)}\rangle ^2}}} \!,
\label{g2t}
\end{equation}
where  
   
	\begin{eqnarray} \label{bb solution}
	&&\! \! \!\!\!\!\!\! \! \!\!\!  \! \!  \langle \delta{{{\hat {\tilde b}}{^\dag} }\!(t)\delta\hat {\tilde b}(t\! + \! \tau )}\rangle \!  =\!  \frac{1}{4} \! \left[ {\frac{{2\mathcal{C}{{\bar n}_m} - 2{s_\varepsilon }}}{{1 \! +\!  2{s_\varepsilon }}}{e^{ {\lambda _1}\tau }}} \right.\left. {\! + \frac{{2\mathcal{C}{{\bar n}_m} \! +\! 2{s_\varepsilon }}}{{1 - 2{s_e}}}{e^{ {\lambda _2}\tau }})}\! \right] \nonumber  \\
	&& \! \! \!\!\!\!\!\!\!\!\! \!  \! \! + \! \frac{{(1 \! +\! 2{{\bar n}_m})}}{4} \! \left[ {\frac{\!1\!-\! \mathcal{C}\! +\! {2{s_\varepsilon }}}{{1\! +\!  2{s_\varepsilon }}}{e^{ {\lambda _1}(2t + \tau)}}} \right.\left. {\!\! +\!  \frac{1\!-\! \mathcal{C} \! - \! {2{s_\varepsilon }}}{{1 - 2{s_e}}}{e^{ {\lambda _2}(2t + \tau)}}} \! \right] \! \! , \\
	&&\! \! \!\!\!\!\!\! \!\!\! \!  \! \! \langle \delta{{{\hat {\tilde b}}{^\dag} }\!(t)\delta{{\hat {\tilde b}}{^\dag} }(t \! + \! \tau )} \rangle \!  = \!  \frac{1}{4} \! \left[ {\frac{{2\mathcal{C}{{\bar n}_m} \!  - \! 2{s_\varepsilon }}}{{1 \! + \! 2{s_\varepsilon }}}{e^{ {\lambda _1}\tau }}} \right.\left. { \! - \frac{{2\mathcal{C}{{\bar n}_m} \! +\! 2{s_\varepsilon }}}{{1 - 2{s_\varepsilon}}}{e^{ {\lambda _2}\tau }})} \! \right] \nonumber \\ 
	&&\! \! \!\!\!\!\!\! \!\!\!\!  \! \! + \! \frac{{(1\!  + \! 2{{\bar n}_m})}}{4} \! \left[ \! {\frac{1\!-\!\mathcal{C}\! +\! {2{s_\varepsilon }}}{{1 \! + \! 2{s_\varepsilon }}}{e^{ {\lambda _1}(2t + \tau)}}} \! \right.\left. {\!\! - \frac{1\! - \! \mathcal{C}\! -\! {2{s_\varepsilon }}}{{1 - 2{s_\varepsilon}}}{e^{ {\lambda _2}(2t + \tau)}}} \! \right] \!\! .
	\end{eqnarray}

\begin{figure}
	\includegraphics[width=8.7cm,height=3cm]{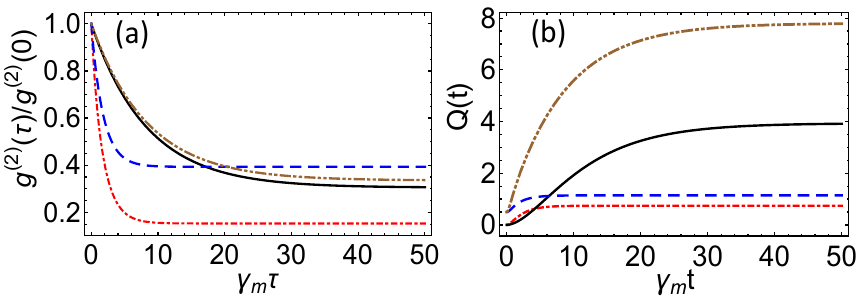}
	\caption{( Color online) (a) Normalized autocorrelation function, ${g^{(2)}}(\tau )/{g^{(2)}}(0)$, and (b) the Mandel parameter, $Q(t)$, obtained analytically within the RWA versus the scaled times $\gamma_m \tau$ and $\gamma_m t$. Black solid line and brown double-dot-dashed line correspond to $s_\varepsilon=0.44$ and initial mean phonon number $ \bar n_m \approx 0.01 $ and $\bar n_m \approx 0.5  $, respectively. Red dot-dashed line and blue dashed line correspond to  $s_\varepsilon=0.22$ and initial mean phonon number $ \bar n_m \approx 0.01 $ and $\bar n_m \approx 0.5  $, respectively.}
	\label{Fig4}
\end{figure}
In Fig.~\ref{Fig4}(a), we have plotted ${g^{(2)}}(\tau )/{g^{(2)}}(0)$ as a function of $\gamma_m \tau$ for two positive values of $s_\varepsilon$ (red-detuned regime) and different values of initial mean phonon number around the ground state of the MO. We find that the generated Casimir phonons exhibit bunching phenomenon which is inhibited as the control parameter $s_\varepsilon$ is increased. The manifestation of phonon bunching is attributed to the possibility of emission of two phonons simultaneously by the phononic DCE, which is a phonon analog of the OPA process described by the Hamiltonain of Eq. (\ref{H_eff_squeezing_final_RWA}). As can be seen by increasing the initial mean phonon number, the phonon bunching decreases rapidly.

To determine the phonon counting statistics, one can calculate the Mandel parameter as follows
\begin{eqnarray} \label{Q(t)}
&&  Q(t) =( \langle \delta{{{\hat {\tilde b}}{^{\dag 2}}}\delta{{\hat {\tilde b}}{^2}}}\rangle - \langle \delta{{{\hat {\tilde b}}{^\dag} }\delta\hat {\tilde b}}\rangle^2) / \langle \delta{{{\hat {\tilde b}}{^\dag} }\delta\hat {\tilde b}}\rangle .
\end{eqnarray}  
For $Q>0$ ($Q<0$), the statistics is super-Poissonian (sub-Poissonian); $Q=0$ stands for Poissonian statistics. To calculate the Mandel parameter, we note that the operator $\delta \hat {\tilde b}$ can be decomposed into two parts: $\delta \hat {\tilde b}(t) = \delta\hat {\tilde b}_d(t) +\delta\hat {\tilde b}_s(t)$ where
\begin{eqnarray}
&& \! \! \! \! \!\!\!\!\!\!\!\!\!\!\!\!\!\!\!\!  \delta \hat  {\tilde b}_d(t)\!\! =\! \delta \hat {\tilde X}(0){e^{{\lambda _1}t}} + i \delta \hat {\tilde P}(0){e^{{\lambda _2}t}} ,\\
&& \! \! \! \! \!\!\!\!\!\!\!\!\!\!\!\!\!\!\!\!  \delta \hat {\tilde b}_s(t) \! \! = \! {e^{{\lambda _1}t}}\! \! \int_0^t \!\!{dt'} \! {e^{ - {\lambda _1}t'}}\!{{\hat {\tilde X}}_{in}}(t') \! + \!  i  {e^{{\lambda _2}t}} \! \!\! \int_0^t \!\! {dt'} \! {e^{ - {\lambda _2}t'}}\!{{\hat {\tilde P}}_{in}}\! (t') ,
\label{BF}
\end{eqnarray}
characterize the deterministic and stochastic evolutions of the MO, respectively. Therefore, we get
\begin{eqnarray}
&&\!\!\!\!\! \!\!\!\!\! \langle \delta{{\hat {\tilde b}^{\dag 2}}\delta{{\hat {\tilde b}}{^2}}}\rangle  = \langle \delta{{\hat {\tilde b}_d^{\dag 2}}(t)\delta{\hat {\tilde b}_d^2}(t)} \rangle \!  + 2{\langle \delta{{\hat{\tilde b}_s^\dag }(t)\delta\hat {\tilde b}_s(t)} \rangle }{^2}\! +\!{\vert {\langle \delta{{\hat {\tilde b}_s^2}(t)} \rangle } \vert }{^2} \nonumber \\
&& \!\!\!\!\! \!\!\!\!\! +  2{\mathop{\rm Re}\nolimits} \langle \delta{{\hat {\tilde b}_d^2}(t)} \rangle \! \langle \delta{{\hat {\tilde b}_s^{\dag 2}}(t)} \rangle \!  + \! 4 \langle \delta{{\hat {\tilde b}_d^\dag }(t)\delta \hat {\tilde b}_d(t)} \rangle \!  \langle \delta{{\hat {\tilde b}_s^\dag }(t)\delta \hat{\tilde b}_s(t)}\rangle ,
\label{bd2b2}
\end{eqnarray}
where 
\begin{subequations} \label{bd2b2simple}
	\begin{eqnarray}
	&& \langle \delta{{\hat {\tilde b}_s^2}} \rangle \! = \frac{{{\gamma _m}}}{4}\mathcal{C}(1 + 2{{\bar n}_m})\left[ {\frac{{{e^{2{\lambda _1}t}} - 1}}{{2{\lambda _1}}} - \frac{{{e^{2{\lambda _2}t}} - 1}}{{2{\lambda _2}}}} \right] \nonumber \\
	&& \qquad\qquad + \frac{i}{2}(\mathcal{C} - 1)(1 + 2{{\bar n}_m})({e^{ - {\gamma _m}t}} - 1), \\
	&& \langle \delta{{\hat {\tilde b}_s^\dag } \delta\hat {\tilde b}_s}\rangle  = \frac{{{\gamma _m}}}{4}\mathcal{C}(1 + 2{{\bar n}_m})\left[ {\frac{{{e^{2{\lambda _1}t}} - 1}}{{2{\lambda _1}}} + \frac{{{e^{2{\lambda _2}t}} - 1}}{{2{\lambda _2}}}} \right] \nonumber \\
	&& \qquad\qquad + \frac{1}{2}({e^{ - {\gamma _m}t}} - 1),  \\
	&& \langle \delta{{\hat {\tilde b}_d^{\dag 2}}\delta{\hat {\tilde b}_d^2}} \rangle  = {R_1}({e^{4{\lambda _1}t}} + {e^{4{\lambda _2}t}}) + {R_2}{e^{ - {\gamma _m}t}}({e^{2{\lambda _1}t}} + {e^{2{\lambda _2}t}}) \nonumber  \\
	&& \qquad\qquad + {R_3}{e^{ - 2{\gamma _m}t}} ,  \\
	&& \langle \delta{{\hat {\tilde b}_d^\dag }\delta\hat {\tilde b}_d} \rangle  = \frac{{(1 + 2{{\bar n}_m})}}{4}({e^{2{\lambda _1}t}} + {e^{2{\lambda _2}t}}) - \frac{1}{2}{e^{ - {\gamma _m}t}}, \\
	&& \langle \delta{{\hat {\tilde b}_d^2}} \rangle  = \frac{{(1 + 2{{\bar n}_m})}}{4}({e^{2{\lambda _1}t}} - {e^{2{\lambda _2}t}}), 
	\end{eqnarray}
\end{subequations}
with ${R_1} = (3 + 6 \langle n \rangle_{\rm th} + 6\langle n^2 \rangle_{\rm th})/16$, ${R_2} =  - (1 + 2\langle n \rangle_{\rm th})/2$, ${R_3}  = (10 + 4\langle n \rangle_{\rm th} + 4\langle n^2 \rangle_{\rm th})/16$ where $ \langle n \rangle_{\rm th}=\bar n_m $ and $ \langle n^2 \rangle_{\rm th}=2 \bar n_m^2 +\bar n_m $.
Figure \ref{Fig4}(b) illustrates the Mandel parameter for the generated Casimir phonons as a function of $\gamma_m t$ for two positive values of $s_\varepsilon$ (red-detuned regime) and different values of $ \bar n_m $. As is seen, the generated Casimir phonons exhibit super-Poissonian statistics. With increasing the parameter $s_\varepsilon$ (for instance, through increasing the power of driving laser or amplitude of modulation) the Mandel parameter increases (enhancement of super-Poissonian statistics), and it is finally stabilized at an asymptotic value. Moreover, the rate in reaching the asymptotic value is inversely proportional to $s_\varepsilon$; the smaller $s_\varepsilon$ is, the more rapidly the Mandel parameter tends to the asymptotic value. Furthermore, super-Poissonian statistics enhances by increasing the initial mean phonon number for the same value of $ s_\varepsilon $.  From Eqs. (\ref{g2t})-(\ref{bd2b2simple}) one can easily check that both the autocorrelation function and the Mandel parameter are invariant under the replacements $s_\varepsilon \! \to \! - s_\varepsilon$ and $\lambda_1 \leftrightarrow \lambda_2$. This means that the counting statistics of the generated Casimir phonons is the same in both red-and blue- detuned regimes.  Moreover, numerical calculations of the autocorrelation function and the Mandel parameter beyond the RWA (not shown here) reveal that the CRTs do not lead to any significant effect on the temporal behaviour of the phonon statistics.

We point out that the counting statistics of the generated phonons can be experimentally readout based on the method introduced in Ref. \cite{Phonon counting measurement}. Photons from an external pump laser are scattered by phonons of the vibrating membrane. Sending the cavity output through a series of narrowband optical filters centered on the cavity resonance suppresses the pump so that photon counting events will correspond directly to counting phonon emission or absorption events. Furthermore, detecting the filtered optical cavity output in a Hanbury-Brown and Twiss setup to measure the second-order photon correlation function will provide a direct readout of the second-order correlation function of the generated phonons.

\section{ Displacement spectrum of the MO} \label{sec4}

We next proceed to calculate the steady-state displacement spectrum of the MO. We show that monitoring the linewidth of the displacement spectrum of the MO provides a way to identify the generation of the Casimir phonons. The symmetrized displacement spectrum of the MO is defined by \cite{Genes}
\begin{equation}
\!\!\!\!{ S_{ xx}} \! (\omega ) \! =\! \frac{1}{4 \pi}\! \! \int \!\! \!  {d\Omega  \langle \delta\hat X(\omega)\delta\hat X(\Omega) \! + \! \delta\hat X(\Omega) \delta\hat X(\omega)  \rangle} {e^{i(\omega +  \Omega )t }} .
\label{spectrum}
\end{equation}
By Fourier transforming Eqs. \ref{X and P equation}(a) and \ref{X and P equation}(b) into the frequency domain, one gets
\begin{subequations} \label{XP_omega}
	\begin{eqnarray}
	&& \delta \hat X(\omega)= \chi_m(\omega) \left[ \chi_\varepsilon \delta \hat P(\omega) + \hat X_{in}(\omega) \right], \\
	&& \delta \hat P(\omega)= \chi_m(\omega) \left[ \chi_\varepsilon \delta \hat X(\omega) + \hat P_{in}(\omega) \right], 
	\end{eqnarray}
\end{subequations} 
where $\chi_m(\omega)= 1/(\gamma_m/2-i\omega)$ is the mechanical susceptibility. By solving the set of algebraic equation \ref{XP_omega}(a) and \ref{XP_omega}(b) one can find the displacement of the MO in terms of input noises as follows
\begin{equation}
\delta \hat X(\omega)= G_{ x}(\omega) \hat X_{in}(\omega) + G_{p}(\omega) \hat P_{in} (\omega) ,
\label{x_omega}
\end{equation}
where $G_{p}(\omega)= \chi_\varepsilon \chi_m(\omega) G_{ x}(\omega)$ and $G_{ x}(\omega)= \chi_m(\omega) / \left( 1-\chi_\varepsilon^2 \chi_m^2(\omega) \right)$. Using the correlation functions in the Fourier space, i.e., $\langle \hat X_{in}(\omega) \hat X_{in}(\Omega) \rangle=\gamma_m/4 (1+2 \bar n_m) \delta(\omega + \Omega)$, $ \langle \hat P_{in}(\omega) \hat P_{in}(\Omega) \rangle=\gamma_m/4 \left( (1+ 2 \bar n_m) + 4\xi^2 \right) \delta(\omega + \Omega) $ and $\langle \hat X_{in}(\omega) \hat P_{in}(\Omega) \rangle=- \langle \hat P_{in}(\omega) \hat X_{in}(\Omega) \rangle=i/2 \delta(\omega + \Omega) $ together with the relations $\chi_m(-\omega)=\chi_m^\ast(\omega)$ and $G_j(-\omega)=G_j^\ast (\omega)$ ($j= x,p$), the symmetrized displacement spectrum of the MO will be obtained as
\begin{eqnarray}
S_{xx}\!(\tilde \omega) \!  = \! \frac{1}{D(\tilde \omega)} \! \! \left[ \! (1 \! + \! 2\bar n_m) (  \! 1 \! + \! 4 {\tilde \omega}^2 )  \! + \!  4s_\varepsilon^2 \! \! \left( \! ( \! 1 \! + \!  2\bar n_m) \! + \! 4\xi^2 \! \right)\! \right]\!,
\label{S_final}
\end{eqnarray}
where $D(\tilde \omega)=16 \gamma_m \left[ \tilde {\omega}^2 +\left( (1-4s_\varepsilon^2)/4 - \tilde{\omega}^2 \right)^2 \right]$ and $\tilde \omega= (\omega-\Omega_m) /\gamma_m$. There are two contributions in Eq. (\ref{S_final}); the first term in the bracket represents the contribution of the input noise quadrature $\hat X_{in}$, while the second term accounts for the contribution of the input noise $\hat P_{in}$. In the absence of time modulation of the driving laser ($\varepsilon=0$) only the contribution of the input noise quadrature $\hat X_{in}$ will survive and the displacement spectrum of Eq. (\ref{S_final}) reduces to ${\bar S_{ xx}} {\vert_{\varepsilon=0} } (\tilde \omega)  =\left(1+2 \bar n_m/\gamma_m\right) \left[ 1/(1+4 \tilde \omega^2) \right]$ which is a Lorentzian with width $\gamma_m$ centered at $\omega=\Omega_m$. This can also be understood from the equations of motion \ref{X and P equation}(a) and \ref{X and P equation}(b) which are decoupled when the modulation is switched off and consequently the time evolution of the mechanical quadrature $\delta \hat X$ is simply given by $\delta \hat X(t)= e^{- \gamma_m/2 t} \left( \delta \hat X_0 + \int_0^t \hat X_{in} (t') e^{\gamma_m t'}  \right)$.

\begin{figure} [h!]
	\includegraphics[width=8.5cm, height=3.5cm]{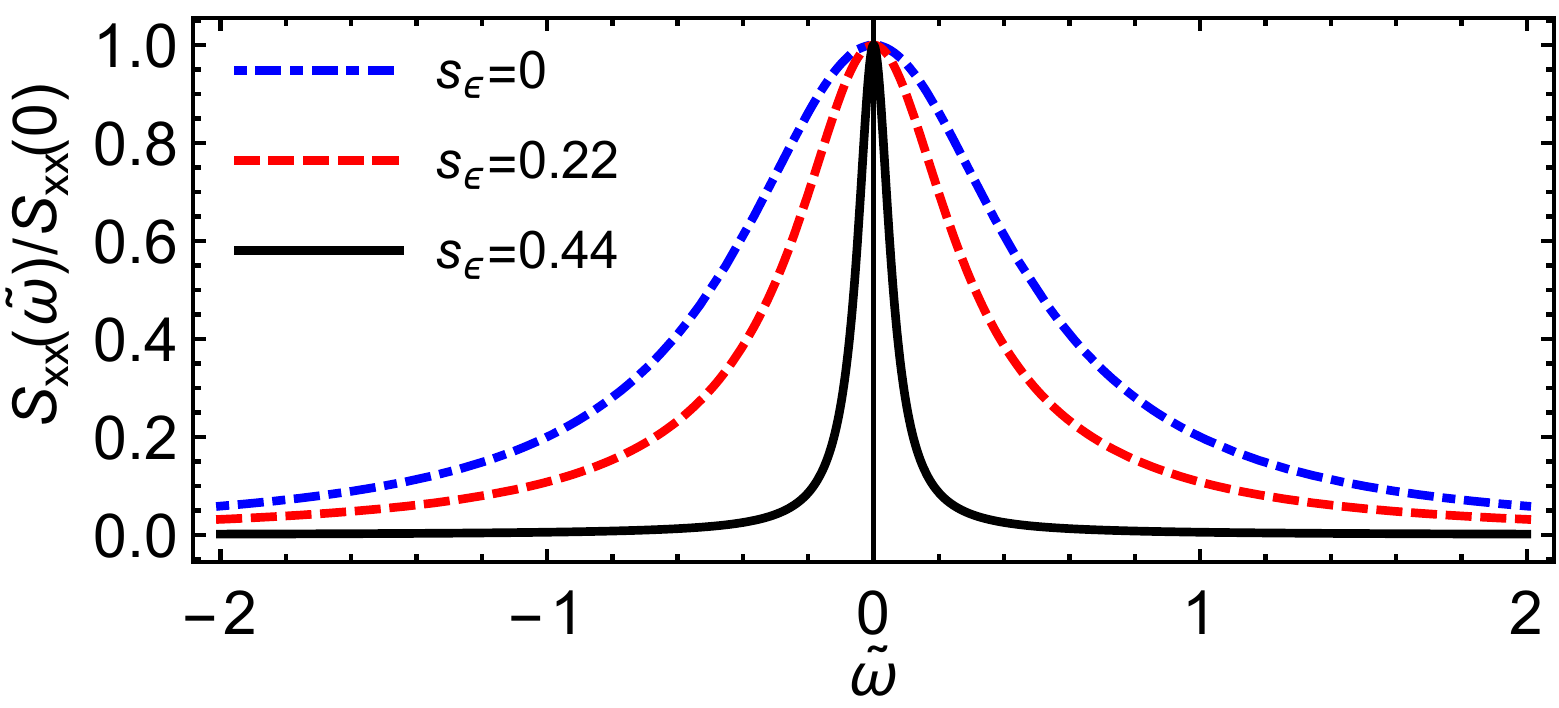}
	\caption{( Color online)  Normalized symmetric displacement spectrum of the MO, obtained analytically within the RWA, versus $\tilde \omega=(\omega-\Omega_m)/\gamma_m$ for two positive values of $s_\varepsilon=0.22$ (red dashed line) and $s_\varepsilon=0.44$ (black solid line) corresponding to the red-detuned regime. The spectrum for the case $s_\varepsilon=0$ (blue dot-dashed line) is plotted for comparison.}
	\label{Fig5}
\end{figure}

To examine how the displacement spectrum of the MO is affected by the time modulation of the driving laser, we have plotted in  Fig. (\ref{Fig5}) the normalized spectrum $S_{xx}(\tilde{\omega})/S_{xx}(0)$ versus the dimensionless frequency $\tilde{\omega}$ for two positive values of $s_\varepsilon$ (red-detuned regime) as well as for the case of no modulation ($\varepsilon=0$). As can be seen, with increasing the value of $s_\varepsilon$ (e.g., through varying the driving laser parameters) the width of the spectrum gets narrower while its peak remains unaltered. In other words, with increasing the value of the amplitude of modulation the quadrature squeezing increases, and therefore the width of the spectrum becomes narrower. Since the creation of the Casimir phonons is attributed to the amplitude modulation of the driving laser, the comparison of the width of the displacement spectrum of the MO in the presence of the time-modulated pumping with one that is obtained by constant pumping provides a way to identify the generation of the Casimir phonons. Moreover, as is evident from Eq. (\ref{S_final}), the displacement spectrum of the MO is invariant under the replacement $s_\varepsilon \to - s_\varepsilon$, indicating that the shape of the spectrum is the same for both red- and blue- detuned
regimes. The only difference is that the peak of the spectrum in the red detuning and blue detuning cases is centered at
$\omega=\Omega_{Casimir}^{red}=\omega_m - 2 \vert \chi_0 \vert$ and $\omega=\Omega_{Casimir}^{blue}=\omega_m + 2\vert \chi_0 \vert$, respectively. Moreover, numerical calculation of the displacement spectrum beyond the RWA (not shown here) reveals that the contribution of the CRTs leads only to a negligible shift of the peak position of the spectrum.

\section{CONCLUSIONS} \label{summary}

In summary, we have shown that the coherent modulation of the driving laser in the far-detuned regime of cavity optomechanics leads to the parametric amplification of the mechanical vacuum fluctuations of the MO or Casimir phonon generation over time scales longer than the cavity lifetime. The generated Casimir phonons exhibit bunching, super-Poissonian statistics and quadrature squeezing which are controllable by the parameters of the driving laser as the external control parameters. The maximum available mechanical noise reduction is about $50 \%$ which is limited by the temperature of the phonon bath such that for temperatures higher than a critical value, there will no longer be squeezing in the MO. The critical temperature depends on the optomechanical and the modulation parameters. We have shown how by controlling the optomechancial parameters one can control the noise induced to the MO by the cavity mode.
Our suggested scheme for controllable generation of the Casimir phonons in a membrane-in-the-middle optomechanical system is realisable in the dispersive regime for  experimentally feasible parameters of the system. We have also found that the time modulation of the driving laser leads to the linewidth narrowing of the displacement spectrum of the MO which in turn is a signature of the generation of the Casimir phonons.

\section*{Acknowledgements}
	The authors are grateful to David Vitali for useful comments and discussions.


\begin{thebibliography}{66}
	\bibitem{Moore} G. T. Moore, ``Quantum Theory of the Electromagnetic Field in a Variable-Length One-Dimensional Cavity," \href{https://doi.org/10.1063/1.1665432}{J. Math. Phys. \textbf{11}, 2679 (1970)}.
	\bibitem{Davies} P. C. W. Davies and S. A. Fulling, ``Radiation from Moving Mirrors and from Black Holes,"  \href{https://doi.org/10.1098/rspa.1977.0130}{Proc. Roy. Soc. London. A \textbf{356}, 237 (1977)} .
	\bibitem{Yablonovitch} E. Yablonovitch, ``Accelerating reference frame for electromagnetic waves in a rapidly growing plasma: Unruh-Davies-Fulling-DeWitt radiation and the nonadiabatic Casimir effect,"  \href{https://doi.org/10.1103/PhysRevLett.62.1742}{Phys.Rev. Lett. \textbf{62}, 1742 (1989)} .
	\bibitem{Dodonov1} V. V. Dodonov, ``Current status of the dynamical Casimir effect,"  \href{https://doi.org/10.1088/0031-8949/82/03/038105}{Phys. Scripta \textbf{82}, 038105 (2010)}.
	\bibitem{Nation} P. D. Nation, J. R. Johansson, M. P. Blencowe, and F. Nori, ``Stimulating uncertainty: Amplifying the quantum vacuum with superconducting circuits," \href{https://doi.org/10.1103/RevModPhys.84.1}{Rev. Mod. Phys. \textbf{84}, 1 (2012)}.
	\bibitem{Schwinger} J. Schwinger, ``Casimir energy for dielectrics," \href{https://doi.org/10.1073/pnas.89.9.4091}{Proc. Natl Acad. Sci. USA \textbf{89}, 4091 (1992)} .
	\bibitem{Wilson} J. R. Johansson, G. Johansson. C. M. Wilson, and F. Nori, ``Dynamical Casimir effect in superconducting microwave circuits," \href{https://doi.org/10.1103/PhysRevA.82.052509}{ Phys. Rev. A \textbf{82}, 052509 (2010)}.
	\bibitem{Davies book} N. D. Birrell and P. C. D. Davies, \textit{Quantum Fields in Curved Space} \href{https://doi.org/10.1017/CBO9780511622632}{(Cambridge University, 1982)} .
	\bibitem{Dodonov2} V. V. Dodonov, ``Dynamical Casimir effect in a nondegenerate cavity with losses and detuning," \href{https://doi.org/10.1103/PhysRevA.58.4147}{Phys. Rev. A \textbf{58}, 4147 (1998)}.
	\bibitem{Dalvit} D. A. R Dalvit and F. D. Mazzitelli, ``Creation of photons in an oscillating cavity with two moving mirrors," \href{https://doi.org/10.1103/PhysRevA.59.3049}{Phys. Rev. A \textbf{59}, 3049 (1999)}.
	\bibitem{Dodonov3} V. V. Dodonov, ``Nonstationary Casimir Effect and Analytical Solutions for Quantum Fields in Cavities with Moving Boundaries," \href{https://doi.org/10.1002/0471231479.ch7}{Adv. Chem. Phys. \textbf{119}, 309 (2001)}.
	\bibitem{Crocce} M. Crocce, D. A. R. Dalvit, and F. D. Mazzitelli, ``Quantum electromagnetic field in a three-dimensional oscillating cavity," \href{https://doi.org/10.1103/PhysRevA.66.033811}{Phys. Rev. A \textbf{66}, 033811 (2002)}.
	\bibitem{Dodonov4} A. V. Dodonov, E. V. Dodonov, and V. V. Dodonov, ``Photon generation from vacuum in nondegenerate cavities with regular and random periodic displacements of boundaries," \href{https://doi.org/10.1016/j.physleta.2003.08.065}{Phys. Lett. A \textbf{317}, 378 (2003)}.
	\bibitem{Dodonov5} A. V. Dodonov, R. Lo Nardo, R. Migliore, A. Messina, and V. V. Dodonov, ``Analytical and numerical analysis of the atom–field dynamics in non-stationary cavity QED," \href{https://doi.org/10.1088/0953-4075/44/22/225502}{J. Phys. B: At. Mol. Opt. Phys. \textbf{44}, 225502 (2011)}.
	\bibitem{Dodonov6} V. V. Dodonov, A. B. Klimov, and V. I. Manko, ``Generation of squeezed states in a resonator with a moving wall," \href{https://doi.org/10.1016/0375-9601(90)90333-J}{Phys. Lett. A \textbf{149}, 225 (1990)}.
	\bibitem{Dodonov7} V. V. Dodonov and M. A. Andreata, ``Squeezing and photon distribution in a vibrating cavity," \href{https://doi.org/10.1088/0305-4470/32/39/301}{J. Phys. A \textbf{32}, 6711 (1999)}.
	\bibitem{Johansson} J. R. Johansson, G. Johansson, C. M. Wilson, P. Delsing, and F. Nori, ``Nonclassical microwave radiation from the dynamical Casimir effect," \href{https://doi.org/10.1103/PhysRevA.87.043804}{Phys. Rev. A \textbf{87},043804 (2013)}.
	\bibitem{Bhattacherjee} N. Aggarwal, A. B. Bhattacherjee, A. Banerjee, and M. Mohan, ``Influence of periodically modulated cavity field on the generation of atomic-squeezed states," \href{https://doi.org/10.1088/0953-4075/48/11/115501}{J. Phys. B: At. Mol. Opt. Phys.\textbf{48}, 115501 (2015)}.
	\bibitem{Felicetti} Felicetti, M. Sanz, L. Lamata, G. Romero, G. Johansson, P. Delsing, and E. Solano, ``Dynamical Casimir Effect Entangles Artificial Atoms," \href{https://doi.org/10.1103/PhysRevLett.113.093602}{Phys. Rev. Lett. \textbf{113}, 093602 (2014)}.
	\bibitem{Sabin} C. Sabin and G. Adesso, ``Generation of quantum steering and interferometric power in the dynamical Casimir effect," \href{https://doi.org/10.1103/PhysRevA.92.042107}{Phys. Rev. A \textbf{92}, 042107 (2015)}.
	\bibitem{Lombardi} C. Braggio, G. Bressi, G. Carugno, C. Del Noce, G. Galeazzi, A. Lombardi, A. Palmieri, G. Ruoso, and D. Zanello, ``A novel experimental approach for the detection of the dynamical Casimir effect," \href{https://doi.org/10.1209/epl/i2005-10048-8}{Europhys. Lett. \textbf{70}, 754(2005)}.
	\bibitem{Dodonov8} V. V. Dodonov and A. V. Dodonov, ``QED effects in a cavity with a time-dependent thin semiconductor slab excited by laser pulses," \href{https://doi.org/10.1088/0953-4075/39/15/S20}{J. Phys. B: At. Mol. Opt. Phys. \textbf{39}, S749 (2006)}.
	\bibitem{Dodonov9} V. V. Dodonov and A. V. Dodonov, ``The nonstationary Casimir effect in a cavity with periodical time-dependent conductivity of a semiconductor mirror," \href{https://doi.org/10.1088/0305-4470/39/21/S18}{J. Phys. A: Math. Gen. \textbf{39}, 6271(2006)}.
	\bibitem{Dezael} F. X. Dezael and A. Lambrecht, ``Analogue Casimir radiation using an optical parametric oscillator," \href{https://doi.org/10.1209/0295-5075/89/14001}{Europhys. Lett. \textbf{89}, 14001 (2010)}.
	\bibitem{Faccio} D. Faccio and I. Carusotto, ``Dynamical Casimir Effect in optically modulated cavities," \href{https://doi.org/10.1209/0295-5075/96/24006}{Europhys. Lett. \textbf{96}, 24006 (2011)}.
	\bibitem{Motazedifard DCE} A. Motazedifard, M. H. Naderi, and R. Roknizadeh, ``Analogue model for controllable Casimir radiation in a nonlinear cavity with amplitude-modulated pumping: generation and quantum statistical properties," \href{https://doi.org/10.1364/JOSAB.32.001555}{J. Opt. Soc. Am. B \textbf{32}, 1555(2015)}.
	\bibitem{Agnesi} A. Agnesi, C. Braggio, G. Bressi, G. Carugno, F. Della Valle, G. Galeazzi, G. Messineo, F. Pirzio, G. Reali, G. Ruoso, D. Scarpa, and D. Zanello, ``MIR: An experiment for the measurement of the dynamical Casimir effect," \href{https://doi.org/10.1088/1742-6596/161/1/012028}{J. Phys.: Conf. Series \textbf{161}, 012028 (2009)}.
	\bibitem{Kawakubo} T. Kawakubo and K. Yamamoto, ``Photon creation in a resonant cavity with a nonstationary plasma mirror and its detection with Rydberg atoms," \href{https://doi.org/10.1103/PhysRevA.83.013819}{Phys. Rev. A \textbf{83}, 013819 (2011)}.
	\bibitem{Pourkabirian} C. M. Wilson, G. Johansson, A. Pourkabirian, M. Simoen, J. R. Johansson, T. Duty, F. Nori, and P. Delsing, ``Observation of the dynamical Casimir effect in a superconducting circuit," \href{https://doi.org/10.1038/nature10561}{Nature(London) \textbf{479}, 376 (2011)}.
	\bibitem{Lahteenmaki} P. , G. S. Lahteenmaki, Paraoanu, J. Hassel, and P. J. Hakonen, ``Dynamical Casimir effect in a Josephson metamaterial," \href{https://doi.org/10.1073/pnas.1212705110}{Proc. Natl. Acad. Sci. U.S.A. \textbf{110}, 4234 (2013)}.
	\bibitem{Recati} I. Carusotto, R. Balbinot, A. Fabbri, and A. Recati, ``Density correlations and analog dynamical Casimir emission of Bogoliubov phonons in modulated atomic Bose-Einstein condensates," \href{https://doi.org/10.1140/epjd/e2009-00314-3}{Eur. Phys. J. D \textbf{56}, 391 (2010)}.
	\bibitem{Jaskula} J. C. Jaskula, G. B. Partridge, M. Bonneau, R. Lopes, J. Ruaudel, D. Boiron, and C. I. Westbrook, ``Acoustic Analog to the Dynamical Casimir Effect in a Bose-Einstein Condensate," \href{https://doi.org/10.1103/PhysRevLett.109.220401}{Phys. Rev. Lett. \textbf{109}, 220401 (2012)}.
	\bibitem{Koghee} S. Koghee and M. Wouters, ``Dynamical Casimir Emission from Polariton Condensates," \href{https://doi.org/10.1103/PhysRevLett.112.036406}{Phys. Rev. Lett. \textbf{112}, 036406 (2014)}.
	\bibitem{Busch} X. Busch, I. Carusotto, and R. Parentani, ``Spectrum and entanglement of phonons in quantum fluids of light," \href{https://doi.org/10.1103/PhysRevA.89.043819}{Phys. Rev. A \textbf{89}, 043819(2014)}.
	\bibitem{Saito} H. Saito and H. Hyuga, ``Dynamical Casimir effect for magnons in a spinor Bose-Einstein condensate," \href{https://doi.org/10.1103/PhysRevA.78.033605}{Phys. Rev. A \textbf{78}, 033605 (2008)}.
	\bibitem{Dodonov10} V. V. Dodonov and J. T. Mendonca, ``Dynamical Casimir effect in ultra-cold matter with a time-dependent effective charge," \href{https://doi.org/10.1088/0031-8949/2014/T160/014008}{Phys. Scr. T \textbf{160}, 014008 (2014)}.
	\bibitem{Mahajan} S. Mahajan, N. Aggarwal, T. Kumar, A. B. Bhattacherjee, and M. Mohan, ``Dynamics of an optomechanical resonator containing a quantum well induced by periodic modulation of cavity field and external laser beam," \href{https://doi.org/10.1139/cjp-2014-0255}{Can. J. Phys. \textbf{93},716 (2015)}.
	\bibitem{Thompson} J. D. Thompson, B. M. Zwickl, A. M. Jayich, F. Marquardt, S. M. Girvin, and J. G. E. Harris, ``Strong dispersive coupling of a high-finesse cavity to a micromechanical membrane," \href{https://doi.org/10.1038/nature06715}{Nature (London) \textbf{452}, 72 (2008)}.
	\bibitem{Jayich} A. M. Jayich, J. C. Sankey, B. M. Zwickl, C. Yang, J. D. Thompson, S. M. Girvin, A. A. Clerk, F.
	Marquardt, and J. G. E. Harris, ``Dispersive optomechanics: a membrane inside a cavity," \href{http://iopscience.iop.org/1367-2630/10/9/095008}{New J . Phys. \textbf{10}, 095008 (2008)}.
	
	
	
	\bibitem{Mari1} A. Mari and J. Eisert, ``Gently Modulating Optomechanical Systems," \href{https://doi.org/10.1103/PhysRevLett.103.213603}{Phys. Rev. Lett. \textbf{103}, 213603 (2009)}.
	\bibitem{Giovannetti} A. Farace and V. Giovannetti, ``Enhancing quantum effects via periodic modulations in optomechanical systems," \href{https://doi.org/10.1103/PhysRevA.86.013820}{Phys. Rev. A \textbf{86}, 013820 (2012)}.
	\bibitem{Mari2} A. Mari and J. Eisert, ``Opto- and electro-mechanical entanglement improved by modulation,"  \href{https://doi.org/10.1088/1367-2630/14/7/075014}{New J. Phys. \textbf{14}, 075014 (2012)}.
	
	\bibitem{two-phonon driving clerk} B. A. Levitan, A. Metelmann, and A. A. Clerk, ``Optomechanics with two-phonon driving,"  \href{https://doi:10.1088/1367-2630/18/9/093014}{New J. Phys. \textbf{18}, 093014 (2016)}.
	\bibitem{PRL modulation} A. Pontin, M. Bonaldi, A. Borrielli, L. Marconi, F. Marino, G. Pandraud, G. A. Prodi,
	P. M. Sarro, E. Serra, and F. Marin, ``Dynamical Two-Mode Squeezing of Thermal Fluctuations in a Cavity Optomechanical System," \href{https://doi.org/10.1103/PhysRevLett.116.103601}{Phys. Rev. Lett. \textbf{116}, 103601 (2016)}.
	\bibitem{singlecavitymoderegime} C. K. Law, ``Interaction between a moving mirror and radiation pressure: A Hamiltonian formulation," \href{https://doi.org/10.1103/PhysRevA.51.2537}{Phys. Rev. A \textbf{51}, 2537 (1995)}.
	\bibitem{singlemechanicalmoderegime} C. Genes, D. Vitali, and P. Tombesi, ``Simultaneous cooling and entanglement of mechanical modes of a micromirror in an optical cavity," \href{https://doi.org/10.1088/1367-2630/10/9/095009}{New J. Phys. \textbf{10}, 095009 (2008)}.
	\bibitem{Sh. Barzanjeh} Sh. Barzanjeh, M. H. Naderi,  and M. Soltanolkotabi, ``Generation of motional nonlinear coherent states and their superpositions via an intensity-dependent coupling of a cavity field to a micromechanical membrane," \href{https://doi.org/10.1088/0953-4075/44/10/105504}{J. Phys. B: At. Mol. Opt. Phys. \textbf{44}, 105504 (2011)}.
	\bibitem{Soltanolkotabi} Sh. Barzanjeh,  M. H. Naderi, and M. Soltanolkotabi, ``Back-action ground-state cooling of a micromechanical membrane via intensity-dependent interaction," \href{https://doi.org/10.1103/PhysRevA.84.023803}{Phys. Rev. A \textbf{84}, 023803 (2011)}.
	
	
	\bibitem{LIGO} V. Braginsky, S. E. Strigin, and S. P. Vyatchanin, ``Analysis of parametric oscillatory instability in power recycled LIGO interferometer," \href{https://doi.org/10.1016/S0375-9601(02)01357-9}{Phys. Lett. A \textbf{305}, 111 (2002)}.
	\bibitem{xsensing1} T. J. Kippenberg and K. J. Vahala, ``Cavity Opto-Mechanics," \href{https://doi.org/10.1364/OE.15.017172}{Opt. Exp. \textbf{15}, 17172 (2007)}.
	\bibitem{CQNCPRL} M. Tsang and C.M. Caves, ``Coherent Quantum-Noise Cancellation for Optomechanical Sensors," \href{https://doi.org/10.1103/PhysRevLett.105.123601}{Phys. Rev. Lett. \textbf{105} 123601 (2010)}.
	\bibitem {CQNCPRX} M. Tsang and C.M. Caves, ``Evading Quantum Mechanics: Engineering a Classical Subsystem within a Quantum Environment," \href{https://doi.org/10.1103/PhysRevX.2.031016}{Phys. Rev. X \textbf{2} 031016 (2012).}.
	\bibitem{CQNCmeystre} F. Bariani, H. Seok, S. Singh, M. Vengalattore, and P. Meystre, ``Atom-based coherent quantum-noise cancellation in optomechanics," \href{https://doi.org/10.1103/PhysRevA.92.043817}{Phys. Rev. A \textbf{92}, 043817 (2015)}.
	\bibitem{CQNCmaximilian} M. H. Wimmer, D. Steinmeyer, K. Hammerer, and M. Heurs, ``Coherent cancellation of backaction noise in optomechanical force measurements," \href{https://doi.org/10.1103/PhysRevA.89.053836}{Phys. Rev. A \textbf{89}, 053836 (2014)}.
	\bibitem{aliNJP} Ali Motazedifard, F. Bemani, M. H. Naderi, R. Roknizadeh and D. Vitali, ``Force sensing based on coherent quantum noise cancellation in a hybrid optomechanical cavity with squeezed-vacuum injection,"  \href{http://stacks.iop.org/1367-2630/18/i=7/a=073040}{New J. Phys. \textbf{18}, 073040 (2016) }.
	\bibitem{ground state cooling} A. D. O’ Connell, M. Hofheinz, M. Ansmann, R. C. Bialczak, M. Lenander, E. Lucero, M. Neeley, D. Sank, H. Wang, M. Weides, J. Wenner, J. M. Martinis, and A. N. Cleland, ``Quantum ground state and single-phonon control of a mechanical resonator," \href{https://doi.org/10.1038/nature08967}{Nature (London) \textbf{464}, 697 (2010)}.
	\bibitem{Sideband cooling} J. D. Teufel, T. Donner, D. Li, J. W. Harlow, M. S. Allman, K. Cicak, A. J. Sirois, J. D. Whittaker, K. W. Lehnert, and R. W. Simmonds, ``Sideband cooling of micromechanical motion to the quantum ground state," \href{https://doi.org/10.1038/nature10261}{Nature (London) \textbf{475}, 359 (2011)}.
	\bibitem{Laser cooling} J. Chan, T. P. M. Alegre, A. H. Safavi-Naeini, J. T. Hill, A. Krause, S. Groblacher, M. Aspelmeyer, and O. Painter, ``Laser cooling of a nanomechanical oscillator into its quantum ground state," \href{https://doi.org/10.1038/nature10461}{Nature (London) \textbf{478}, 89 (2011)}.
	\bibitem{Borkje}A. Nunnenkamp, K. Borkje, and S. M. Girvin, ``Single-Photon Optomechanics," \href{https://doi.org/10.1103/PhysRevLett.107.063602}{Phys. Rev. Lett. \textbf{107}, 063602 (2011)}.
	\bibitem{Hammerer} K. Jähne, C. Genes, K. Hammerer, M. Wallquist, E. S. Polzik, and P. Zoller, ``Cavity-assisted squeezing of a mechanical oscillator," \href{https://doi.org/10.1103/PhysRevA.79.063819}{Phys. Rev A. \textbf{79}, 063819 (2009)}.
	
	
	\bibitem{Tombesi} C. Genes, A. Mari, D. Vitali, and P. Tombesi, ``Quantum effects in optomechanical systems," \href{http://dx.doi.org/10.1016/S1049-250X(09)57002-4}{Adv. At. Mol. Phys. \textbf{57}, 33 (2009)}. 
	\bibitem{Gardiner} C. W. Gardiner and P. Zoller, \textit{Quantum Noise} \href{https://doi.org/10.1007/978-3-662-04103-1}{(Springer, Berlin, 2000)}.
	
	 \bibitem{Thomson} J. D. Thompson, B. M. Zwickl, A. M. Jayich, F. Marquardt, S. M. Girvin, and J. G. E. Harris, ``Strong dispersive coupling of a high-finesse cavity to a micromechanical membrane," \href{https://doi:10.1038/nature06715}{Nature (London) \textbf{452}, 72 (2008)}.
	
	
	\bibitem{Genes} C. Genes, D. Vitali, P. Tombesi, S. Gigan, and M. Aspelmeyer, ``Ground-state cooling of a micromechanical oscillator: Comparing cold damping and cavity-assisted cooling schemes," \href{https://doi.org/10.1103/PhysRevA.77.033804}{Phys. Rev. A \textbf{77}, 033804 (2008)}.
	
	\bibitem{Aspelmeyer} M. Aspelmeyer, T. J. Kippenberg, and F. Marquardt, “ Cavity optomechanics," \href{https://doi.org/10.1103/RevModPhys.86.1391}{Rev. Mod. Phys. \textbf{86}, 1391 (2014)}. 
		
 
	
	\bibitem{Ruth-Hurwitz} E. X. DeJesus and C. Kaufman, ``Routh-Hurwitz criterion in the examination of eigenvalues of a system of nonlinear ordinary differential equations," \href{https://doi.org/10.1103/PhysRevA.35.5288}{Phys. Rev. A. \textbf{35}, 103605 (2013)}.
	
    
   \bibitem{marquartcooling}  F. Marquardt, J. P. Chen, A. A. Clerk, S. M. Girvin, ``Quantum theory of cavity-assisted sideband cooling of mechanical motion," \href{https://doi.org/10.1103/PhysRevLett.99.093902}{Phys. Rev. Lett. \textbf{99}, 093902 (2007)}. 
    
   \bibitem{Vitali.statistic measurement} D. Vitali, S. Gigan, A. Ferreira, H. R. Bohm, P. Tombesi, A. Guerreiro, V. Vedral, A. Zeilinger, and M. Aspelmeyer, “Optomechanical entanglement between a movable mirror and a cavity field,” 
   \href{https://doi.org/10.1103/PhysRevLett.98.030405}{Phys. Rev. Lett. \textbf{98}, 030405 (2007)}.
   
    
	\bibitem{Walls} D. F. Walls and G. J. Milburn, \textit{Quantum Optics} \href{https://doi.org/10.1007/978-3-642-79504-6}{(Springer-Verlag,1994)}.
	
	
	\bibitem{Phonon counting measurement} J. D. Cohen, S.M. Meenehan, G. S. MacCabe, S. Groblacher, A. H. Safavi-Naeini, F. Marsili, M. D. Shaw, and O. Painter, “Phonon counting and intensity interferometry of a nanomechanical resonator,” \href{https://doi.org/10.1038/nature14349}{Nature \textbf{520}, 522 (2015)}.
	
\end{thebibliography}
\end{document}